\documentstyle[11pt,epsfig,aaspp4]{article}
\tighten 
\begin{document}

\newcommand{\lya}{Lyman~$\alpha$}
\newcommand{\lyb}{Lyman~$\beta$}
\newcommand{\za}{$z_{\rm abs}$}
\newcommand{\ze}{$z_{\rm em}$}
\newcommand{\cmtwo}{cm$^{-2}$}
\newcommand{\nhi}{$N$(H$^0$)}
\newcommand{\degpoint}{\mbox{$^\circ\mskip-7.0mu.\,$}}
\newcommand{\Ha}{\mbox{H$\alpha$}}
\newcommand{\Hb}{\mbox{H$\beta$}}
\newcommand{\hgamma}{\mbox{H$\gamma$}}
\newcommand{\kms}{\,km~s$^{-1}$}      
\newcommand{\minpoint}{\mbox{$'\mskip-4.7mu.\mskip0.8mu$}}
\newcommand{\mv}{\mbox{$m_{_V}$}}
\newcommand{\Mv}{\mbox{$M_{_V}$}}
\newcommand{\peryr}{\mbox{$\>\rm yr^{-1}$}}
\newcommand{\secpoint}{\mbox{$''\mskip-7.6mu.\,$}}
\newcommand{\sqdeg}{\mbox{${\rm deg}^2$}}
\newcommand{\squig}{\sim\!\!}
\newcommand{\subsun}{\mbox{$_{\twelvesy\odot}$}}
\newcommand{\et}{et al.~}

\def\ltsima{$\; \buildrel < \over \sim \;$}
\def\simlt{\lower.5ex\hbox{\ltsima}}
\def\gtsima{$\; \buildrel > \over \sim \;$}
\def\simgt{\lower.5ex\hbox{\gtsima}}
\def\arcs{$''~$}
\def\arcm{$'~$}
\vspace*{0.1cm}
\title{
THE REST FRAME OPTICAL SPECTRA OF LYMAN BREAK GALAXIES:
STAR FORMATION, EXTINCTION, ABUNDANCES, AND KINEMATICS\altaffilmark{1}}

\vspace{1cm}
\author{\sc Max Pettini}
\affil{Institute of Astronomy, Madingley Road, Cambridge, CB3 0HA, UK}
\author{\sc Alice E. Shapley and Charles C. Steidel\altaffilmark{2}}
\affil{Palomar Observatory, California Institute of Technology,
MS~105--24, Pasadena, CA 91125}
\author{\sc Jean-Gabriel Cuby}
\affil{European Southern Observatory, Alonso de Cordova 3107,
Santiago, Chile}
\author{\sc Mark Dickinson}
\affil{Space Telescope Science Institute, 3700 San Martin Drive,
Baltimore, MD 21218}
\author{\sc Alan F.M. Moorwood}
\affil{European Southern Observatory, Karl-Schwarzschild-Str. 2, 
D-85748 Garching, Germany}
\author{\sc Kurt L. Adelberger\altaffilmark{3}}
\affil{Harvard-Smithsonian Center for Astrophysics, 60 Garden Street,
Cambridge, MA 02138}
\author{\sc Mauro Giavalisco}
\affil{Space Telescope Science Institute, 3700 San Martin Drive,
Baltimore, MD 21218}

\altaffiltext{1}{Based on data obtained at the European
Southern Observatory on Paranal, Chile, and at the
W.M. Keck Observatory on Mauna Kea, Hawaii.
The W.M. Keck Observatory is operated as a scientific partnership 
among the California Institute of Technology, the University of 
California, and NASA, and was made possible by 
the generous financial support of the 
W.M. Keck Foundation.}
\altaffiltext{2}{Packard Fellow.}
\altaffiltext{3}{Harvard Junior Fellow.}

\newpage
\begin{abstract}
We present the first results of a spectroscopic survey
of Lyman break galaxies (LBGs) in the near-infrared 
aimed at detecting the emission lines of [O~II], [O~III],
and \Hb\ from the H~II regions of normal star forming galaxies at $z 
\simeq 3$. From observations of 15 objects with 
the Keck and VLT telescopes, augmented with data from the literature
for an additional four, we reach the following main conclusions.
The rest-frame optical properties of LBGs at the bright
end of the luminosity function are remarkably uniform; 
their spectra are dominated by emission lines,
[O~III] is always stronger than \Hb\ and [O~II],
and projected velocity dispersions are 
between 50 and 115\,km~s$^{-1}$.
Contrary to expectations, the star formation rates deduced
from the \Hb\ luminosity are on average no larger than those
implied by the stellar continuum at 1500\AA; 
presumably any differential 
extinction between rest-frame optical and UV
wavelengths is small compared with the relative uncertainties
in the calibrations of these two star formation tracers.
For the galaxies in our sample,
the abundance of oxygen can only be determined to within one 
order of magnitude without recourse to other emission lines
([N~II] and \Ha) which are generally not available.
Even so, it seems well established that LBGs are
the most metal-enriched structures at $z \simeq 3$, apart from QSOs,
with abundances greater than about 1/10 solar and generally
higher than those of damped \lya\ systems at the same epoch.
They are also significantly overluminous for their
metallicities; this is probably an indication that their
mass-to-light ratios are low 
compared with present-day galaxies.
At face value the measured velocity dispersions  
imply virial masses of about $10^{10}\,M_{\odot}$
within half-light radii of 2.5\,kpc.
The corresponding mass-to-light ratios, 
$M/L \approx 0.15$ in solar units, 
are indicative of stellar populations
with ages between $10^8$ and $10^9$ years,
consistent with the optical-IR spectral energy distributions.
However, we are unable to establish conclusively 
if the widths of the emission lines do reflect 
the motions of the H~II regions
within the gravitational 
potential of the galaxies, even though in two cases
we see hints of rotation curves.
All 19 LBGs observed show evidence for galactic-scale superwinds;
such outflows have important consequences for regulating star formation, 
distributing metals over large volumes, and allowing Lyman continuum
photons to escape and ionize the intergalactic medium.
\end{abstract}

\newpage
\section{Introduction}

Our current knowledge of the normal galaxy population at high
redshift relies heavily on observations in the rest frame
ultraviolet (UV) region, where we see the integrated light of OB
stars and the strongest spectral features are interstellar
absorption lines (e.g. Steidel et al. 1996, 1999; Lowenthal et
al. 1996; Pettini et al. 2000; Steidel, Pettini, \& Adelberger
2001). The situation is quite different at low redshift where the UV
spectra of star forming galaxies have become accessible only
relatively recently, requiring observations from space with the
{\it Hubble Space Telescope\/}¥ ({\it HST\/}¥; see Leitherer 2000 for a
recent overview). For the last fifty years local galaxies with
active star formation have been studied first and foremost at 
optical wavelengths through the rich emission line spectrum
produced by their H~II regions.

There is therefore a strong incentive to obtain such spectra for
galaxies at $z \simeq 3$ where the samples of UV-selected objects
now number nearly one thousand. Apart from the surprises often
associated with opening a new wavelength window, there are
several obvious scientific motivations. The luminosity of the
Balmer lines, primarily \Ha\ and \Hb, gives a measure of the star
formation rate which is directly comparable to the values deduced
in local surveys. Furthermore, since the optical emission lines
and the far-UV continuum do not respond to dust extinction to the
same degree, the relative luminosity of a galaxy in these two
tracers of star formation could in principle be used as a
reddening indicator. When integrated over an entire galaxy, the
widths of the nebular lines should reflect the velocity
dispersion of the H~II regions within the overall gravitational
potential so that a kinematical mass may be deduced. This is not
possible in the UV, because the interstellar absorption lines are
sensitive to small amounts of gas accelerated to high velocities
by energetic events such as supernova explosions and bulk
outflows, while the stellar absorption lines from OB stars are
intrinsically broad. Finally, there are well established chemical
abundance diagnostics based on the relative strengths of nebular
emission lines, primarily [O II], [O III] and \Hb. Abundance
measurements are much more difficult in the UV where the more
easily observed interstellar lines are generally saturated so
that their equivalent widths depend mostly on  the velocity
dispersion of the gas and only to a lesser extent on the column
density of the absorbing ions.

At $z \simeq 3$ the most prominent lines from H~II regions are
redshifted into the near-infrared (IR) where observations are
more challenging than at optical wavelengths. Not
only does atmospheric absorption by water vapour limit
observations to specific wavelength bands, but even in these the
night sky glows brightly in numerous OH emission lines which vary
in intensity over timescales of minutes. However, with a large
sample of galaxies such as that produced by the Lyman break
technique (Steidel 2000) it is possible to isolate redshifts which
place the transitions of interest in gaps between the OH sky
lines; here the background is sufficiently dark for faint
extragalactic work to become possible. In Pettini et al.  (1998)
we proved the feasibility of this approach with $K$-band
observations of five Lyman break galaxies (LBGs) using the CGS4
spectrograph on the 3.8~m United Kingdom Infrared Telescope
(UKIRT). Although successful, these initial attempts showed that
galaxies at $z \simeq 3$ are generally too faint to be studied in
large numbers with 4-m class telescopes. The five galaxies
targeted by Pettini et al. are among the brightest LBGs known and
still required essentially one night of integration per object
with CGS4. On the other hand, the recent commissioning of
near-IR, high resolution spectrographs on the Keck~II and VLT1
telescopes now makes it possible to extend this work to a larger
and more representative sample of high redshift galaxies. We have
begun such a program of near-IR spectroscopy and present here our
results for 15 Lyman break galaxies. Preliminary reports have
been published in conference proceedings (Pettini 2000; Cuby et
al. 2000) and similar observations of individual galaxies have
also been obtained by Teplitz et al. (2000a,b) and Kobulnicky \&
Koo (2000).

This paper is arranged as follows. In \S2 we summarize details of
the observation and data reduction procedures. Individual objects
with features of special interest are discussed in \S3. In \S4 we
present our measurements of \Hb\ luminosity, derive star
formation rates, and compare them with the values deduced from
the UV continuum. \S5 deals with the abundance of Oxygen derived
from the strong line ratios, while in \S6 we analyze the
kinematics of the galaxies observed. The most important results
of this work are reviewed in \S7. Unless otherwise stated,
we adopt a $\Omega_{\rm M} = 0.3$, $\Omega_{\Lambda} = 0.7$,
$H_0 = 70$\kms\ Mpc$^{-1}$ cosmology throughout the paper.\\

\section{Observations}

Details of the Lyman break galaxies observed are collected in
Table 1. The optical redshifts listed in columns (4) and (5) were
measured from 12.5~\AA\ resolution spectra obtained with the Low
Resolution Imaging Spectrograph on the Keck telescopes. 
The optical photometry is from our $U_n$, $G$, ${\cal R}$ deep
imaging at the prime foci of the Palomar 5-m, 
La Palma 4.2-m, and Kitt Peak 4-m
telescopes, while the (${\cal R} - K_{\rm AB}$) colors listed in
column (7) are from the survey by Shapley et al. (in preparation)
conducted with the Keck Near-Infrared Camera (NIRC). Objects were
selected from a variety of fields, some including high redshift
QSOs and radio galaxies, others being blank regions of sky
already targeted by other deep surveys such as the Hawaii deep
field SSA22, the Caltech Deep Field, and the Groth-Westphal
strip. Coordinates of all the objects observed are given in 
columns (2) and (3) of Table 1. 

The primary selection criterion
in compiling the sample in Table 1 was redshift; as explained
above we aimed to minimize the chance of interference from bright
sky lines. We also limited ourselves to galaxies brighter than
${\cal R} \simeq 24.2$ which at $z \simeq 3$ corresponds to 
$\sim 1.3\,L^{\ast}$ in the rest frame ultraviolet
(Steidel et al. 1999);
thus the galaxies observed are drawn from the brightest
10\% of the full sample of more than 800 spectroscopically
confirmed LBGs with ${\cal R} \leq 25.5$ (Steidel 2000).
We further attempted to 
span a range of ($G - {\cal R}$) colors so as to sample objects
with different UV continuum slopes, presumably reflecting
different degrees of dust extinction. 
None of the galaxies show evidence for 
nuclear activity in their rest frame UV spectra.
Most of the observations were targeted at 
\Hb\ ($\lambda 4861.32$) and the
[O~III] doublet ($\lambda\lambda 4958.91, 5006.84$) in the
$K$-band; in a few additional cases we also searched for the
corresponding [O~II] ($\lambda\lambda 3726.05, 3728.80$) in the
$H$-band.

\subsection{Data Acquisition}

The data were secured during four observing runs in 1999 and 2000
using the NIRSPEC and ISAAC spectrographs on the Keck~II and VLT1
telescopes respectively (see last column of Table 1).

A detailed description of NIRSPEC is given by McLean et al.
(1998). It uses a $1024 \times 1024$ 27$\mu$m pixel (ALADDIN2)
InSb detector; in the medium dispersion mode employed for the
present observations each detector pixel corresponds to 0.143
arcsec in the spatial direction and the dispersion in the
spectral direction is 4.2~\AA/pixel. In the September 1999
observing run we used an entrance slit 0.57~arcsec wide which
gives a spectral resolution of 11.6~\AA\ FWHM (resolving power $R
\simeq 1750$ at 2.05~$\mu$m) sampled with 2.8 wavelength bins
(measured from the widths of sky emission lines in the reduced
spectra). In April 2000 we changed to the wider 0.76~arcsec
entrance slit, giving a resolution of 13.6~\AA\ FWHM (resolving
power $R \simeq 1500$) sampled with 3.2 bins. The galaxies to be
observed were placed in the slit by offsetting from a nearby
bright star. This manoeuvre was 'blind' in the sense that
normally the galaxy and the offset star were separated by more
than the 42~arcsec length of the NIRSPEC slit; however, on no
occasion did we fail to acquire the target. Individual exposure
times were 900~s; the detector was read out in multiple read mode
(16 reads at the start and the end of each integration) to reduce
noise. After each exposure the offsetting procedure was repeated
and the target reacquired at a different position on the slit, 
typically 5~arcsec from the previous one. Conditions were mostly
photometric during both Keck runs and the typical seeing in the
$K$-band was 0.5--0.6~arcsec FWHM. 

The short wavelength channel of ISAAC (Moorwood et al. 1999) uses
a  $1024 \times 1024$ 18.5$\mu$m pixel Rockwell array; the
projected pixel scale along the slit is very similar to that of
NIRSPEC, 0.146 arcsec/pixel. The spectral dispersion, however, is
2.3 times higher at 1.24~\AA/pixel; with the 1~arcsec wide slit
we used, the resolution is 7.4~\AA\ FWHM ($R \simeq 2750$)
sampled with six wavelength bins. Acquisition was also achieved
via nearby bright stars; in this case however the 120~arcsec long
ISAAC slit allowed us to have both offset star and target galaxy
on the slit at the same time (having rotated the slit to the
required position angle on the sky). The bright star spectrum is
a useful reference for verifying the accuracy of the offsetting
procedure, and monitoring seeing and sky transparency 
fluctuations. Our two VLT runs in September and November 1999
were mostly photometric and the seeing varied between 0.5 and
0.9~arcsec. The ISAAC observations were performed in
beam-switching mode with the object moved between two positions
on the slit separated by 10~arcsec. After an A-B-B-A series of
$4 \times 900$\,s long exposures, the object was reacquired at a different
position along the slit, and the four-exposure sequence repeated.

\subsection{Data Reduction}

Examples of NIRSPEC and ISAAC spectra of Lyman break galaxies in
the $K$-band are reproduced in Figures 1 and 2 respectively,
while Figure 3 shows detections of [O II]~$\lambda 3727$ in the
$H$-band. In Figure 2 the ISAAC spectra have been rebinned to
twice the original pixel size, so that the spectral resolution is
sampled with three wavelength bins in these plots. 

The two-dimensional data frames were reduced using a series of
IRAF scripts, following the same basic steps for
NIRSPEC and ISAAC spectra. 
After flat-fielding (by reference to the spectrum of a 
quartz halogen lamp), correcting for defective pixels
(flagged from the analysis of the noise
in flat-field and dark current frames), 
and rotating the images
(so as to align the spectrum approximately along the columns of the
detector), we applied a geometric transformation to correct for 
the spatial distortion and spectral curvature introduced by 
the spectrograph optics. 
The two-dimensional map used in this
transformation was obtained by 
stepping a bright star in regular intervals along the slit
and determining the star trace as a function of slit position.
The rectified 2-D images were wavelength calibrated 
on a vacuum scale with the aid of the atlas of  
OH sky emission lines by Rousselot et al. (2000).
The next steps in the reduction differed slightly
between NIRSPEC and ISAAC spectra.

With NIRSPEC we moved the object along the slit
every 900\,s exposure. Consequently, we used the 
exposures immediately preceding and following a 
given image to construct a sky frame which was then subtracted
in 2-D from each image, after appropriate line-by-line scaling.
A further background correction was applied
to each sky-subtracted image (by fitting a polynomial
in the spatial direction at each wavelength bin)
to remove the residuals at the wavelengths of the
strongest sky lines. The individual frames so processed
were then registered and co-added (with a
rejection algorithm to exclude cosmic rays and other 
invalid pixels) to produce a fully reduced, 
two-dimensional spectrum of each galaxy observed.
The final step in the reduction involved extracting
one-dimensional spectra by summing the signal along the slit;
we applied both weighted and unweighted extractions,
but found the two methods to give very similar results.
The IRAF scripts used in the manipulation of the data frames
also generate at each step a corresponding 2-D frame 
of the statistical $1 \sigma$ error appropriate to each pixel;
in the last step this is used to produce 
the one-dimensional error spectrum applicable to 
the extracted galaxy spectrum.

Since the ISAAC observations were obtained in 
beam-switching mode, sky subtraction was achieved
by adding separately the pair of frames with
the object at position A and the pair
with the object at position B. 
Two difference images were then 
produced from these pairs: A$-$B and B$-$A.
After registering one difference image
onto the other by the known A-B telescope offset,
the two were added together; this produces
a characteristic 2-D image with a positive 
galaxy spectrum flanked by two negative spectra
(separated from the positive spectrum by the 
A-B spatial offset). 
Our observations usually consisted of
a series of three or five such A-B-B-A
patterns (at different positions
along the slit); these were combined and extracted
in the same way as the NIRSPEC spectra.

\subsection{Flux Calibration}

The extracted one-dimensional spectra were put on an absolute
flux scale by reference to spectra of bright A0 stars observed
each night at approximately the same airmass as the galaxies
and with the same instrumental configuration. In some cases the
stellar spectra were additionally recorded through a wide slit,
so as to verify empirically the slit losses with the
standard set-ups described at \S2.1.
The stars, which typically have $K$,$H \simeq 7$, were selected
from the list of UKIRT photometric standards\footnote{Available 
at http://www.jach.hawaii.edu/JACpublic/UKIRT/astronomy/calib}
(see also van der Bliek, Manfroid, \& Bouchet 1996).
The conversion factor (as a function of wavelength) between counts
and flux (in Jy units) was based on the spectral energy
distribution of Vega (Colina, Bohlin, \& Castelli 1996)
and the measured broad band $K$ and $H$ magnitudes
of the A stars. 
Since the spectra of A0 stars are relatively smooth at
the wavelengths of interest here, they were also
used to correct for atmospheric absorption (although in only
a few cases did the nebular emission lines fall 
near telluric features).

It is important to assess the overall reliability of the
spectrophotometric calibration---a notoriously difficult step in
the reduction of narrow-slit spectra---because it impacts
directly on the comparison at \S4 below between Balmer line and
UV continuum luminosities, and on the abundance determinations in
\S5. Observations of different A0 stars (on the same night and on
different nights) indicate a random uncertainty of $\leq 15$\%
($1 \sigma$) in the zero point of the flux calibration. 
However, we are more
concerned with systematic errors which may result from inaccurate
placing of the targets in the spectrograph slit and from finite
aperture effects. Generally, we do not expect the latter to be very
large given that the typical half-light radius of Lyman break
galaxies in the rest-frame optical is 0.2--0.4\,arcsec
(Giavalisco, Steidel, \& Macchetto 1996; Dickinson 2000; 
Calzetti \& Giavalisco 2000), but see \S6 below.
We have two ways to check on such
systematics. One is to compare observations of the same object
obtained with different telescopes. Out of the present sample,
Q0201$+$113~C6 was observed in the $K$-band with three different
instruments: NIRSPEC on Keck~II, ISAAC on VLT, and CGS4 on UKIRT.
The three measurements of [O~III]~$\lambda 5007$ at $z = 3.0548$,
respectively ($4.2 \pm 0.4$), ($7.1 \pm 0.9$), and ($4.6 \pm
0.9$) in units of $10^{-17}$~erg~s$^{-1}$~cm$^{-2}$, are within
$\sim 2 \sigma$ of one another.\footnote{As explained in \S3.1
below, this line was misidentified as [O~III]~$\lambda 4959$ at
$z = 3.094$ in the CGS4 spectrum analyzed by Pettini et al.
(1998).} In a separate test, for two objects where a continuum
signal is clearly detected in our NIRSPEC spectra---West~MMD11
(Figure 4) and SSA22a~D3---we can compare it directly with the
broad-band $K$ magnitude (Table 1). In both cases we find that
the spectra underestimate the continuum flux by the same factor
of 1.4\,. We intend to continue monitoring the issue of 
spectrophotometric accuracy as our survey progresses. On the
basis of these limited tests it appears that our
absolute fluxes may be in error by up to about 50\%. We shall
take this uncertainty into account in the analysis of our
results.\\

\section{Comments on Individual Objects}

From Figures 1--3 it can be readily realized that the rest-frame
optical spectra of the Lyman break galaxies in the present sample
share many similarities. The spectra are dominated by the strong
nebular lines, as is the case in local H~II galaxies, and the 
rest-frame blue continuum is generally too weak to be detected.
There is a remarkable uniformity in the line widths (as we shall
see at \S6 below, the one dimensional velocity dispersion is
between $\sigma = 50$ and 115~km~s$^{-1}$), and in the relative
line strengths, with \Hb\ generally weaker, or at most as strong
as, [O~III]$\lambda 4959$. This probably explains why \Hb\ is
undetected in most of our ISAAC spectra which, because of their 
higher dispersion and the smaller aperture of the VLT, generally
have a higher detection threshold than the NIRSPEC spectra. If
these line ratios are typical of most galaxies at $z \simeq 3$,
[O~III]$\lambda 5007$ is then the most prominent spectral feature
to fall in near-IR wavelength region, rather than \Ha. The
generally high excitation of the H~II gas is further reflected by
the weakness of [O~II]$\lambda 3727$ relative to [O~III]$\lambda
5007$ in all five cases observed; compounded with the higher
density of strong sky lines in the $H$-band, this makes  $\lambda
3727$ a difficult feature to detect, as can be appreciated by
comparing Figure 3 with Figures 1 and 2. Thus the prospects for
using [O~II] to probe galaxies at $z > 4$, where [O~III] and \Hb\
are redshifted beyond the $K$-band, are not encouraging even with
8-10~m telescopes.

Before discussing further the properties of the sample as a whole, we
comment briefly on a few objects.

\subsection{Comparison with UKIRT Observations}

We reobserved two galaxies out of the five in the sample by
Pettini et al. (1998), Q0201$+$113~C6 and DSF~2237+116a~C2 (the
latter shown in the bottom panel of Figure 1). With the better
spectra now available we realize that the spectral features in
the former were misidentified, and the emission line redshift
(vacuum heliocentric) is $z_{\rm H~II} = 3.0548$ rather than
3.094 as reported earlier. The confusion arose because we
incorrectly identified [O~III]$\lambda 5007$ as $\lambda 4959$
and a noise feature as \Hb. The new value of redshift is in much
better agreement with that of the interstellar absorption lines
in the rest-frame UV, thereby removing the anomaly noted by
Pettini et al. In the case of DSF~2237+116a~C2, we confirm the
redshift, but not the strengths and widths of the emission lines.
In the CGS4 data this galaxy appeared to have lines twice as
strong and wide as those of the other four objects observed,
although we did express concerns about possible noise
contamination. The values measured from the much higher
signal-to-noise ratio NIRSPEC spectrum, $\sigma =
100$~km~s$^{-1}$, $W_{\rm H\beta} = 25$~\AA\ (rest frame
equivalent width), and $W_{\rm 5007} = 128$~\AA, are smaller by 
about a factor of 2 than those reported in Pettini et al,
bringing this galaxy in line with the rest of the sample.

\subsection{Null Detections}

Out of the 15 LBGs targeted by our survey so far, 13 have yielded
detections of nebular emission lines. The two objects where no
signal could be found in the reduced two-dimensional spectra are
Q1422$+$231~D78 and DSF~2237+116b~C21 (see Table 1). The latter
was observed in poor seeing conditions; furthermore it is the
highest redshift object in the present sample, placing
[O~III]~$\lambda 5007$ at 2.2~$\mu$m where the thermal background
at Paranal is more than five times higher than at 2.0~$\mu$m. The
redshift of the former is uncertain and may have been
misidentified. Another possibility is that these are objects
caught in a post-starburst phase, when the emission lines have
faded but the UV continuum remains bright. From these limited
statistics we conclude that the fraction of bright
LBGs in a post-starburst phase is at most $\sim 13$\%.

\subsection{Q0256$-$000~C17}

In this case we were surprised to find {\it two} emission line
spectra along the slit, one at the position of the LBG and the other displaced
by 3.15~arcsec to the NW (see Figure 14 of Cuby et al. 2000). The
two spectra, reproduced in the middle two panels of Figure 2, are
similar but not identical; in particular the second set of lines
(Q0256$-$000~C17b) is blueshifted by 42\kms\ relative to the
first. In our cosmology the spatial separation between the two
emitting regions
corresponds to a projected distance of 
$24 h_{70}^{-1}$\,kpc. 
Upon re-examination of our broad-band images,
we can see a very faint object at the position of this second [O~III]
emitter with ${\cal R} = 25.97$ (below our magnitude limit
for selecting $U$-drop candidates), ($G - {\cal R}$)\,$= 0.86$,
and ($U_n - G$)\,$> 1.2$\,.
With the assumption that the mean 
$\langle {\cal R} - K_{\rm AB} \rangle = 0.74$ of the present
sample applies to Q0256$-$000~C17b too (see \S4 below),
we deduce a rest
frame equivalent width $W_{5007} = 630$\,\AA. While
relatively rare, H~II galaxies with such a high equivalent width
of [O~III]~$\lambda 5007$ are not unknown locally; for example, in the
catalog by Terlevich et al. (1991) 31 out of 425 emission line
galaxies have $630 \leq W_{5007} \leq 2500$\,\AA.

\subsection{West MMD11}

Among the objects observed is West~MMD11, one of the few LBGs
to have been detected at sub-mm wavelengths with SCUBA
(Chapman et al. 2000). This object has unusually red optical-UV
colors; with ${\cal R} - K_{\rm AB} = 2.72$ it stands out from
the rest of the sample in Table 1 which has 
$\langle {\cal R} - K_{\rm AB} \rangle = 0.74 \pm 0.35$ 
($1 \sigma$). As can be seen
from Figure 4, our NIRSPEC observations clearly show the rest
frame optical continuum, only detected in one other object,
SSA22a~D3. However, it can also be readily appreciated from 
Figure 4 that the emission line spectrum of West~MMD11 is in no way
atypical; the strengths and widths of [O~III] and \Hb, as well as
their ratio, are similar to those of the other galaxies in our
sample (see also Tables 2 and 4, and Figures 5 and 8 below). 
As pointed out by
Adelberger \& Steidel (2000), the unusually red color of
West~MMD11 cannot be entirely attributed to dust extinction, a
conclusion consistent with the findings reported here. Perhaps
the optical continuum is indicative of an earlier strong burst of
star formation, well separated in time from the one which
produced the OB stars we now see in the UV continuum and H~II
region emission lines.

\subsection{Other Objects in Table 1}

When considering the near-IR properties of Lyman break galaxies
in \S4--6 below, we have expanded the present sample of 15 objects
with published observations for an additional four. 
Two, Q0000$-$263~D6 and B2~0902+343~C6, are from our earlier
UKIRT work; although these data have to considered as tentative,
given the discussion at \S3.1 above, we note that the
relevant measurements are typical of the rest of the sample.
Our conclusions would remain unaltered if we excluded these two 
galaxies from the analysis. 
The other two LBGs, West~CC13 and MS~1512-cB58, have been
observed in the $K$-band with NIRSPEC by 
Teplitz et al. (2000a,b respectively)
using a similar instrumental setup as that described 
at \S2.1 above, although the integration times 
were shorter than those of our own observations.
Other relevant details are given in Table 1.\\

\section{Star Formation Rates and Dust Extinction}

Up to now most estimates of the star formation activity in galaxies
at $z \simeq 3$ have been obtained from consideration of their
ultraviolet continuum luminosity. 
In the $K$-band we have access to another 
star formation rate indicator in the \Hb\ line.
Apart from the obvious importance of measuring a physical quantity 
in two independent ways, there are at least two additional incentives
to comparing UV and Balmer line luminosities.
First, many of the surveys at lower redshifts
have been based on \Ha\ (e.g. Gallego et al. 1995; Salzer et al. 2000;
Tresse \& Maddox 1998; Glazebrook et al. 1999;
Yan et al. 1999; Moorwood et al. 2000); 
clearly, it is desirable to use the same measure
(or two closely related ones such as \Ha\ and \Hb)
when considering the evolution with redshift of global properties 
such as the volume averaged star formation rate density.
Second, almost by definition, all interstellar reddening
curves rise from the optical to the ultraviolet.
Thus, at least in principle, it should be possible
to deduce the degree of extinction from the comparison 
between the values of star 
formation rate deduced from the UV continuum 
and from the Balmer lines.

Data relevant to this comparison are collected in Table 2.
In our enlarged sample there are 14 LBGs where we either detect 
the \Hb\ emission line or can place a useful upper limit
to its flux. In columns (6)--(8) of Table 2 we 
list respectively the measured \Hb\ fluxes, 
rest frame equivalent widths, and luminosities.
The values of equivalent width
were derived by referring the measured line fluxes
to the continuum fluxes implied by the 
broad-band $K_{\rm AB}$ magnitudes\footnote{The emission 
lines themselves make a negligible contribution to $K_{\rm AB}$} 
according to the standard relation:
\begin{equation}
AB = -48.60 - 2.5\,{\rm log}\,f_{\nu}
\end{equation}
where $f_{\nu}$ is the flux
in units of erg~s$^{-1}$~cm$^{-2}$~Hz$^{-1}$.
In four cases where a $K$-band magnitude is not available,
we adopted the mean 
$\langle {\cal R} - K_{\rm AB} \rangle = 0.74 \pm 0.35$ 
($1 \sigma$) for the sample (excluding 
West~MMD11 as discussed at \S3.4 above).
The UV continuum luminosities at 1500~\AA\ listed in column (5)
were calculated from the ${\cal R}$ magnitudes
in column (3) also using eq.(1) and applying, 
where necessary, $k$-corrections
deduced from the continuum slope measured in our LRIS spectra. 
Since the effective wavelength of our ${\cal R}$ filter,
6850~\AA, corresponds to a rest frame wavelength of 1670~\AA\ at the 
median $z = 3.1$, the required 
$k$-corrections were normally less than 10\%. 
Finally, columns (9) and (10) of Table 2 give the 
the star formation rates implied by the \Hb\ and UV
continuum luminosities listed in columns (8) and (5) respectively,
adopting Kennicutt's (1998) calibrations:
\begin{equation}
	{\rm SFR (M_{\odot} \, yr^{-1}) = 
	1.4 \times 10^{-28}\,L_{FUV} (erg~s^{-1}~Hz^{-1}) ,}
	\label{}
\end{equation}¥
\begin{equation}
	{\rm SFR (M_{\odot} \, yr^{-1}) = 
	7.9 \times 10^{-42}\,L_{\Ha} (erg~s^{-1}) ,}
	\label{}
\end{equation}¥
\noindent and assuming a ratio \Ha/\Hb\,=\,2.75 (Osterbrock 1989).
Both transformations are appropriate to 
continuous star formation with a Salpeter
initial mass function (IMF) between 0.1 and 100~M$_{\odot}$;
we note that eq. (2) gives values of SFR 40\% higher than 
the calibration we adopted in Pettini et al. (1998).
More generally, both conversions from luminosity
to SFR are subject to significant uncertainties,
as discussed by Kennicutt (1998) and more recently
re-emphasized by Charlot \& Longhetti (2001).
However, here we are concerned mainly with 
{\it comparing} the values of SFR deduced from 
these two indices; since they both trace massive stars
the hope is that many of the systematic uncertainties
may be reduced when considering the ratio
SFR$_{\rm H\beta}$/SFR$_{\rm UV}$.

Nevertheless, several qualifications
are necessary before proceeding further.
First, the conversion factor \Ha/\Hb\,=\,2.75 
does not take into account corrections to the \Hb\ 
flux for reddening and stellar absorption. 
We suspect that both are likely to be of secondary importance, however. 
The values of $E$($B-V$) implied by the slopes
of the UV continua of most Lyman break galaxies
correspond to small Balmer decrements;
for example in MS~1512-cB58, which with 
$E$($B-V$)\,$\simeq 0.3$ is among
the more reddened LBGs, dust extinction 
reduces the \Hb/\Ha\ ratio by only $\sim 10$\%
relative to the recombination value
(Teplitz et al. 2000b).
The integrated \Hb\ stellar absorption line is 
expected to have an equivalent width 
$W_{\rm abs} \simlt 5$~\AA\ (Kobulnicky, Kennicutt, \& Pizagno 1999)
and is therefore likely to result in a small correction 
(comparable to the statistical error) for all the 
entries in Table 2, except for the two cases where
we measure $W_{\rm H\beta} < 10$~\AA.
A potentially more serious effect which may lead us to underestimate
SFR$_{\rm H\beta}$ relative to SFR$_{\rm UV}$ is light loss through the 
spectrograph slit, since the former is 
derived from a spectrophotometric measurement
whereas the latter is from broad-band photometry.
Our limited internal checks described at 
\S2.3 above suggest that such an underestimate may be 
less than $\sim 50$\%, although future observations may 
lead to a revised figure.
 
With these reservations in mind, we compare
the two sets of measurements of SFR in Figure 5,
where the ratio SFR$_{\rm H\beta}$/SFR$_{\rm UV}$
is plotted against the {\it intrinsic}
far-UV spectral slope. The most reliable
measure of the latter, free from the 
uncertainties inherent in narrow slit spectrophotometry,
is the ($G - {\cal R}$) color, corrected
for the opacity of the \lya\ forest according
to the statistical prescription of Madau (1995).
Values of ($G - {\cal R}$)$_{\rm corr}$ are tabulated in 
column (4) of Table 2; 
($G - {\cal R}$)$_{\rm corr} = 0.0$ and $1.0$ correspond
to a UV spectral index $\beta = -2$ and $+0.6$ respectively, 
when the spectrum is approximated by a power law
of the form $F_{\lambda} \propto \lambda^{\beta}$.

It can be readily realized from Figure 5 and Table 2 
that the values of SFR deduced from \Hb\ and from the
far-UV continuum agree within a factor of $\sim 2$ in nearly
every case. There is no tendency for 
SFR$_{\rm H\beta}$ to be systematically greater than SFR$_{\rm UV}$,
nor any evidence for the ratio of these two
quantities to be higher in galaxies with a redder
UV continuum. Earlier suggestions of such a trend
in samples of only a few objects (Pettini et al. 1998;
Teplitz et al. 2000a) now appear to have been the artifact
of small number statistics\footnote{The apparent trend
in Figure 5 of Pettini et al. (1998) was due 
to a single object, DSF~2237$+$116a~C2, whose
\Hb\ flux we have now revised downward by 
a factor of $\sim 3$ (\S3.1).}.

In approximately half of the cases we find 
SFR$_{\rm H\beta} \simlt $\, SFR$_{\rm UV}$.
Apart from the caveats expressed above,
there are several plausible explanations
for this apparently surprising result.
Even without appealing to differences
in the IMF, the Balmer lines are expected to
vary on shorter timescales than the UV continuum
if  `continuous' star formation is in reality 
an approximation to a series of 
individual starburst episodes
(Glazebrook et al. 1999; Bunker, Moustakas, \& Davis 2000).
This is because the overwhelming
contribution to the ionizing flux reprocessed
in the Balmer lines is from the most massive stars,
at the very tip of the IMF, while a wider range
of stellar masses produces
the continuum at 1500~\AA.
In this scenario, LBGs with 
SFR$_{\rm H\beta} < $\, SFR$_{\rm UV}$
are galaxies observed 5--10~Myr after
a major star formation event.
Other possibilities include higher reddening
of the Balmer lines than the UV continuum,
again related to the different evolutionary timescales of the 
stars involved (Calzetti 1997),
and the leakage of some of the Lyman continuum photons 
from the galaxies (Steidel, Pettini \& Adelberger 2001).
In this respect Lyman break galaxies at $z \simeq 3$
are actually not dissimilar to UV selected galaxies
at lower redshifts, where recent comprehensive 
studies have shown that \Ha\ and the UV continuum 
are qualitatively consistent with each other,
although generally at lower luminosities than those
of the present high-$z$ sample (Sullivan et al. 2000;
Bell \& Kennicutt 2001).
In any case, the simplest conclusion 
from the results in Figure 5 is
that, for the Lyman break galaxies in the present sample
(which is representative of the range of ($G - {\cal R}$)
colors of our full survey),
the higher extinction at UV wavelengths
predicted by all reddening curves is evidently masked
by the uncertainties discussed above in relating
UV continuum and \Hb\ luminosities.
The differential extinction between the continuum at 
1500\,\AA\ and \Hb\ must therefore 
be relatively small, as indeed expected on the basis
of the reddening `recipe' by Calzetti (1997).

In concluding this section, we point out that
our results contradict two commonly held
views. First, it is often stated that the UV continuum 
is an unreliable measure of the SFR and that the Balmer lines are to be 
preferred, partly because they suffer less extinction
and partly because they have been studied
in much larger samples of nearby galaxies.
However, adding up the entries in columns (9) and (10)
of Table 2 (and taking the upper limits as detections) 
we find that, for the 14 Lyman break galaxies
considered here, the total values of 
SFR$_{\rm H\beta}$ and SFR$_{\rm UV}$
are within 15\% of each other.
Thus both SFR indicators, {\it uncorrected for dust extinction\/}¥,
give essentially the same star formation rate density.
Secondly, it is felt by some that the relative strengths of the Balmer 
lines give a more secure estimate of the colour excess
$E$($B - V$) than the slope of the far-UV continuum. 
However, as we have seen, in most LBGs at $z \simeq 3$
the Balmer decrement is too small to be measured with the required
precision from ground-based IR spectroscopy, whereas
the UV spectral index $\beta$ can be readily obtained
from broad-band optical photometry.\\

\section{The Oxygen Abundance}

Measurements of element abundances in high redshift galaxies
hold powerful clues to their evolutionary status, to their
links with today's galaxy populations, and to the onset of star 
formation in the universe. Up to now progress on these issues
has relied almost exclusively on studies of damped \lya\
systems (e.g. Pettini et al. 1999; Ellison et al. 2001; 
Prochaska, Gawiser, \& Wolfe 2001 and references therein)
which, however, generally do not probe 
the most active star forming sites 
(e.g. Bunker et al. 1999; Kulkarni et al. 2000, 2001
and references therein).
In contrast, we are still essentially ignorant
of the degree of metal enrichment reached by Lyman break galaxies.
Yet these objects could be important
contributors to the census of metals at high redshifts
which currently seems to show a marked deficit---so far
we can account for only about 1/10 of the element
production associated with the UV light density
at $z \simgt 3$ (Pettini 1998; Pagel 2000).

To date there is only one LBG where element abundances 
have been determined with some degree of confidence, the 
gravitationally lensed galaxy MS~1512-cB58 at $z = 2.7290$,
where young stars, H~II regions, and neutral interstellar
gas all exhibit a metallicity of about 1/3 solar
(Pettini et al. 2000; Teplitz et al. 2000b; Leitherer
et al. 2001). Here we provide measurements for an additional
four LBGs, albeit of lower accuracy than allowed
by the gravitationally lensed nature of MS~1512-cB58.

Approximate estimates of the oxygen abundance in 
H~II regions have traditionally been obtained
from the strong line index $R_{23}$ which relates (O/H)
to the relative intensities of 
[O~II]~$\lambda 3727$, [O~III]~$\lambda\lambda 4959, 5007$,
and \Hb\ (Pagel et al. 1979).
Kobulnicky et al. (1999) have recently re-examined the 
accuracy of this method when applied to the integrated spectra
of distant galaxies and 
in doing so addressed a number of potential problems
including the effects of abundance gradients, inhomogeneous
temperature and ionization, stellar absorption and other complications.
They concluded that, given spectra of sufficient signal-to-noise ratio,
$R_{23}$ typically measures the abundance of oxygen 
to within $\pm 0.2$\,dex.

Measurements in the four galaxies in our sample 
for which we cover all four 
emission lines are collected in Table 3. Values of 
$R_{23} \equiv {\rm (}F_{5007} + F_{4959} + F_{3727}{\rm )}/F_{\rm H\beta}$
are listed in column (9); the range reflects the statistical
$1 \sigma$ errors in the individual values of $F$.
Again, we made no attempts to correct the observed fluxes for
differential extinction because we expect this 
to be a small effect compared
with other uncertainties.
In deriving the oxygen abundances
we made use of the most recent formulation by
Kobulnicky et al. (1999) of the
analytical expressions by McGaugh (1991); 
these formulae express (O/H)
in terms of $R_{23}$ {\it and\/}¥ the ionization index 
$O_{32} \equiv {\rm (}F_{5007} + F_{4959}{\rm )} / F_{3727}$\,.
The results are illustrated in Figure 6 which shows the well-known 
double-valued nature of the relation. 
In the last two columns of Table 3 we 
list separately the values of (O/H) for the lower and upper branches;
in each column the range of values includes the 
statistical uncertainties in both $R_{23}$ and $O_{32}$\,.
We also include in the table the results by Teplitz et al. (2000b)
for MS~1512-cB58.\footnote{We note in passing that, as pointed out most
recently by Pilyugin (2000), the calibration by McGaugh (1991)
overestimates the upper branch value of (O/H)
by about 0.2\,dex when compared with the empirical best fit to
accurate abundances determined 
from temperature sensitive line ratios.
However, the analytical expressions given in Kobulnicky et al. (1999)
partly compensate for this effect.}

It can be seen from these data that 
Lyman break galaxies generally tend to lie 
toward the right-hand side of the diagram
with values log~$R_{23} \simgt 0.7$\,.
In four out of five cases
(the exception being Q0201$+$113~C6)
the statistical errors in the emission line ratios
are tolerable, leading to an uncertainty in 
(O/H) of only about a factor of two. 
Much more serious is the ambiguity
resulting from the double-valued nature of the 
relationship between (O/H) and $R_{23}$.
Only in one case, SSA22a~D3, do we measure a value of 
$R_{23}$ which falls in an unequivocal region
of the diagram and deduce (O/H)\,$ = 8.04$--8.55,
(between 1/6 and 1/2 of solar).
More generally, the two possible solutions
differ by nearly one order of magnitude;
oxygen could be as abundant as in the interstellar 
medium near the Sun, where
(O/H)$_{\rm ISM} = 2/3$\,(O/H)$_{\odot}$
(Meyer, Jura, \& Cardelli 1998; Esteban et al. 1998)
or as low as in low luminosities H~II galaxies
at $\approx 1/10$ solar.

This unsatisfactory state of affairs can in principle be resolved
by observing the [N~II]~$\lambda\lambda 6548, 6584$ and
\Ha\ emission lines; the secondary nature of nitrogen
leads to a strong dependence of the [N~II]/\Ha\ 
intensity ratio on (O/H)
(Kobulnicky et al. 1999; Terlevich, Denicolo, \& Terlevich 2001).
It was on this basis that Teplitz et al. (2000b)
concluded that MS~1512-cB58 lies on the upper branch
of (O/H)~vs.~$R_{23}$ relation, 
and the resulting (O/H)\,$\approx 1/3$\,(O/H)$_{\odot}$
does indeed agree with the abundances of 
other elements measured
in OB stars and H~I gas.

The fact that for two objects (MS~1512-cB58 and SSA~D3)
it has been possible to determine with some degree of confidence
that the oxygen abundance is relatively high
is in our view insufficient ground to assume that this 
is {\it generally\/}¥ the case
in Lyman break galaxies (e.g. Teplitz et al. 2000a;
Kobulnicky \& Koo 2000).
Although we see a great deal of uniformity
in the ratio of [O~III]/\Hb\ among all the LBGs
observed (Figures 1 and 2),
the normally high values of R$_{23}$ implied could still
hide a substantial spread in (O/H), as demonstrated by Figure 6.
The only reliable conclusions we can draw are that
the Lyman break galaxies in our sample:
(a) do not have super-solar abundances
and (b) are significantly more metal-rich than
damped \lya\ systems at the same epoch,
since the latter typically have metallicities
$Z \approx 1/30\,Z_{\odot}$
at $z \simeq 3$ (Pettini et al. 1999; Prochaska et al. 2001).
If either of these two statements were incorrect,
then we would expect log\,$R_{23} \leq 0.5$ which 
can already be excluded by the measured
[O~III]/\Hb\ ratios without recourse to [O~II]~$\lambda 3727$.

The inclusion of [N~II] and \Ha\ in the 
abundance analysis
is generally not an option 
(at least from the ground) for galaxies 
at $z \simeq 3$ 
where the [N~II] doublet and \Ha\ lines
are redshifted beyond the $K$-band.
Even at more favourable redshifts
near $z = 2.3$ for example, 
where all the transitions of interest 
fall within atmospheric transmission windows,
the determination of nebular abundances in
Lyman break galaxies remains a difficult task.
First, there will be relatively few cases 
where the full complement
of [O~II], \Hb\, [O~III], \Ha, and [N~II]
lines is well clear of 
OH sky emission. Second, 
recording all the lines in the three near-IR bands
with sufficient resolution and S/N will
require nearly one night of observations
on an 8-10\,m class telescope for a single LBG.
Thus it appears that, even when multi-object IR spectrographs
become available,
assembling a moderately large sample of
(O/H) measurements at high redshifts
will involve a major observational effort.
There is therefore a strong incentive
to explore alternative abundance indicators
in the rest-frame UV region
which is more easily studied from the ground
(e.g. Leitherer et al. 2001).

\subsection{Metallicity-Luminosity Relation}
 
Present-day galaxies exhibit a clear trend between
$B$-band luminosity and the oxygen abundance 
of their H~II regions (e.g. Skillman, Kennicutt, \& Hodge 1989). 
This metallicity-luminosity relation extends across
morphological types and over 9 magnitudes in luminosity,
and appears to hold at least back to $z \simeq 0.4$ 
(Kobulnicky \& Zaritsky 1998).
Presumably it reflects the fundamental role 
which galaxy mass plays in determining the degree of 
chemical enrichment of the interstellar medium, 
through either the rate at which elements are 
produced by star formation, 
or the ease with which they can escape the 
gravitational potential of the galaxy (or both).
It is clearly of interest to assess whether Lyman break
galaxies conform to this relation.
We address this question in Figure 7, where the low redshift
points are from the compilation by
Kobulnicky \& Koo (2000) after adjustment to our
cosmology (we refer the reader to that paper for 
references to the individual sets of measurements).
The box showing the location of the Lyman break galaxies
encompasses the range of values of (O/H) determined 
for the five LBGs above, 
and the range of rest-frame $B$-band luminosities 
for the {\it full\/}¥ sample (excluding West~MMD11
which is atypical).
Values of $M_{\rm B}$ were deduced from the observed
(or estimated) $K$-band magnitudes and are listed in column 
4 of Table 4 in \S6.1 below.\footnote{Strictly speaking,
at the median $z = 3.1$ the observed $K$-band
corresponds to wavelengths between rest-frame
$B$ and $V$. However, the small $k$-corrections
to rest-frame $B$ (Shapley et al. in preparation)
are unimportant for the purpose of the 
present discussion.}

It is evident from Figure 7 that LBGs at $z \simeq 3$ 
do not conform to today's metallicity-luminosity
relation, as already noted by Kobulnicky \& Koo (2000). 
Even allowing for the  
uncertainties in the determination of (O/H)
discussed above, LBGs fall below
the local line of best fit and have much lower
oxygen abundances than expected for their luminosities.
This is a secure statement; for our objects to fall
on the line, their metallicities would have to be 
well above solar. In this regime 
$F_{\rm H\beta} > F_{\rm [O~III]}$
(log\,$R_{23} < 0$ in Figure 6), as is
the case in present-day nuclear starbursts
(Ho, Filippenko, \& Sargent 1997),
whereas in {\it all\/}¥ our galaxies  
$F_{\rm H\beta} \ll F_{\rm [O~III]}$.
The most obvious interpretation
of this result is that Lyman break galaxies
have mass-to-light ratios which are significantly lower
than those which apply to the normal galaxy
population at the present epoch (see also \S6.1 below).
In this respect, they are more extreme examples
of today's H~II galaxies, which also tend to lie
below the line of best fit in Figure 7.
Another possibility is that the whole
metallicity-luminosity relation is displaced to
lower abundances at high redshifts,
when the universe was younger and the total interval
of time available for the accumulation
of the products of stellar nucleosynthesis was shorter.
It should be possible to determine the 
magnitude of this second effect by measuring the 
oxygen abundance in known samples of galaxies at $z \simeq 1$,
a project which is within the capabilities of current instrumentation.\\

\section{Kinematics}

\subsection{Velocity Dispersions}

The one parameter which is most easily measured from our data
is $\sigma$, the one dimensional velocity dispersion of the H~II gas
along the line of sight, since all that is required
is the detection of one line
(usually [O~III]~$\lambda 5007$). 
Accordingly, this is the physical quantity for which we have
the most extensive set of values, 16 in total.
They are listed in column (7) of Table 4 (after
correction for the instrumental broadening)
and plotted in Figure 8 vs. the far-UV and
$B$-band luminosities (deduced from the 
observed ${\cal R}$ and $K$ magnitudes as 
explained in \S4 and \S5.1 respectively).
It can be seen from the Table and Figure 
that the galaxies in our sample
exhibit a relatively narrow range 
of values of $\sigma$, between 
$\sim 50$ and $\sim 115$\,km~s$^{-1}$.
The median value is $\sigma \simeq 70$\,km~s$^{-1}$.

In Pettini et al. (1998) we used the velocity
dispersions of the nebular lines to estimate the masses of the 
LBGs observed; the enlarged data set available now
essentially confirms the conclusion of that work
that $M > 10^{10} M_{\odot}$.
In the idealized case of a sphere
of uniform density,
\begin{equation}
	M_{\rm vir} = 5 \times \sigma^2 \, r_{1/2}/G
	\label{}
\end{equation}¥
or 
\begin{equation}
	M_{\rm vir} = 1.2 \times 10^{10} M_{\odot}~ 
	\frac{\sigma^2}{100\,{\rm km~s^{-1}}}~
	\frac{r_{1/2}}{{\rm kpc}}
	\label{}
\end{equation}¥
\noindent where $G$ is the gravitational constant and 
$r_{1/2}$ is the half-light radius.
{\it HST} images of Lyman break galaxies
obtained with WFPC2 and NICMOS show that 
they typically have 
$r_{1/2} = 0.2$--0.4\,arcsec,
irrespectively of whether
they are observed in the rest-frame UV or optical light
(Giavalisco et al. 1996; 
Dickinson 2000; Calzetti \& Giavalisco 2000).
In our cosmology, $r_{1/2} = 0.3$\,arcsec
corresponds to $2.3 h_{70}^{-1}$\,kpc and the median
$\sigma = 70$\,km~s$^{-1}$ therefore translates to 
$M_{\rm vir} \simeq 1.3 \times 10^{10} h_{70}^{-1} M_{\odot}$.

A similar value would result for a disk-like geometry.
Extensive simulations by Rix et al. (1997), including
random orientation and other effects,
have shown that $\sigma \approx 0.6 \times V_{\rm c}$,
where $V_{\rm c}$ is the maximum circular velocity
of the ionized gas. Since presumably $V_{\rm c}$ 
is reached at radii 
$r \geq r_{1/2}$, the enclosed dynamical mass
$M_{\rm dyn} =  V_{\rm c}^2 \times r/G$ 
is not very different from the value of $M_{\rm vir}$
obtained from eq.(5)\,.
It is likely, however, that these
masses do not reflect the whole gravitational 
potential of the galaxies but rather refer mainly to
the central, high surface brightness regions.
Locally, it is found that 
in nuclear starbursts and blue compact galaxies
the optical emission lines do not span the full extent
of the rotation curve
traced by the 21~cm line of H~I (Lehnert \& Heckman 1996;
Pisano et al. 2001).

When combined with the median $M_{\rm B} = -22.60$
of our sample (uncorrected for extinction),
the above mass estimate
$M \simeq 1.3 \times 10^{10}\,M_{\odot}$
implies mass-to-light ratios $M/L \approx 0.15$
in solar units. Not surprisingly, this value is  
much lower than those measured in the inner regions 
of galaxies today, which are in the range 2~--~10
(e.g. Binney \& Tremaine 1987). Note that 
this increase in luminosity for a given mass
by a factor of $\approx 30$---or 3.7 magnitudes 
in $M_{\rm B}$---is similar to the 
horizontal offset of LBGs
from today's metallicity-luminosity relation
shown in Figure 7. Thus, our galaxies
seem to have mass-to-light ratios 
typical of young stellar populations. 
From the {\it Starburst99\/}¥ models of 
Leitherer et al. (1999), adjusted to 
a lower mass limit $M_{\rm low} = 0.1\,M_{\odot}$,
we find that $M/L = 0.15$ is intermediate
between the values appropriate to continuous star formation
lasting for $10^8$ ($M/L = 0.05$) and $10^9$ ($M/L = 0.23$) years.
This result is in good agreement with the conclusion
by Shapley et al. (in preparation) that 
the optical to near-IR spectral energy distributions
of most LBGs in their sample are indeed indicative 
of ages between $10^8$ and $10^9$ years.

Nevertheless, it remains to be established
whether in Lyman break galaxies at $z = 3$
the line widths we measure do reflect mostly the overall
velocity dispersion among different star-forming 
regions---the underlying assumption
to using them as tracers of mass---as opposed to
being dominated by
outflows and other large-scale 
motions of a non-gravitational origin
(see \S6.3 below). 
It is evident from Figure 8
that there no correlation between velocity
dispersion and either UV or optical luminosity.
While it could be argued that any such trend
would be difficult to discern given the small sample size and the 
narrow range of absolute magnitudes probed,
we are struck by the results of Adelberger et al.
(in preparation).  
Their much larger (several hundred)
sample of Balmer break galaxies at $z \simeq 1$
not only exhibits values of $\sigma$ between
$\sim 50$ and $\sim 100$\,km~s$^{-1}$, 
similar to those found here,
but also shows no correlation between velocity
dispersion and luminosity over 
a range of nearly five magnitudes.\footnote{This result
is not necessarily in conflict with the work
of Vogt et al. (1997) who found the Tully-Fisher
relation to hold, with only mild evolution, out to
$z \sim 1$. These authors specifically targeted
galaxies with disk morphologies for their 
study, while the Balmer break galaxies 
selected by Adelberger et al.
are a much more heterogeneous sample.}
Thus, until the physical origin
of the broadening of the nebular lines
in Lyman break galaxies is clarified.
it would be unwise to draw far-reaching conclusions
from the mass estimates derived above.

\subsection{Spatially Resolved Line Profiles}

We can try and address this last question 
with ISAAC observations of two of our objects,
Q0347$-$383~C5 and SSA22a~MD46, where we 
unexpectedly discovered spatially resolved 
emission along the spectrograph slit.
As can be seen from Figures 9 and 10,
the [O~III] doublet lines are tilted in 
the 2-D images of these galaxies, hinting at a regular
pattern of velocities as may be produced by a rotating disk.
A third such case (out of a sample of six objects)
has recently been reported by Moorwood et al. (2000)
from \Ha\ observations of a galaxy at $z = 2.192$.  
However, these authors found larger velocity
spreads---and consequently deduced larger 
masses---than determined here, as we now discuss.

To investigate the kinematics of the 
ionized gas, we extracted the spectra
in intervals of two spatial increments,
each 0.146 arcsec on the sky. In this way we optimally
sampled the seeing profile which was measured
to have FWHM = 0.62 and 0.64 arcsec on the
final stacked 2-D images
obtained with integration times of 18\,000 and 14,400\,s
for Q0347$-$383~C5 and SSA22a~MD46 respectively.\footnote{Recall
that all the ISAAC observations included a bright star
on the spectrograph slit; this stellar
spectrum conveniently provides a  
measure of the seeing and an astrometric reference point.}
At the redshifts of these two galaxies,
$z = 3.2337$ and 3.0855 respectively, 
we sample the spatial structure of the emission
at projected intervals on the sky
of $2.2 h_{70}^{-1}$\,kpc in our cosmology.

The bottom right-hand panels in Figures 9 and 10
show the runs of relative velocities along the slit,
measured from the central wavelengths
of [O~III]~$\lambda 5007$ by Gaussian fitting.
The velocity ranges spanned by the line {\it centres\/}¥
are small, only about $\pm 30$ and $\pm 40$\,km~s$^{-1}$
in Q0347$-$383~C5 and SSA22a~MD46 respectively,
over linear projected distances of 
$\pm 4.5$\,kpc. Taken at face value,
these measurements would imply
much smaller masses than 
derived at \S6.1 above,
$M_{\rm dyn} \sim 1.3 \times 10^9 h_{70}^{-1}\,M_{\odot}$.
However, in reality this value is a very conservative lower
limit to the enclosed mass of ionized gas because in each case:
(a) the line profiles are only just resolved spatially,
and the whole `rotation curve' on either side of the
center is sampled with only two points\footnote{It is easy 
to see that, as each sample point is an  
average over regions with different 
velocities in a rising rotation curve, the net effect
is an underestimate of $V_{\rm c}$ since the 
the emission line intensity decreases from the center
of the galaxy.};
(b) we do not know the inclination angle
of the galaxy; and (c) we do not know
the orientation of the spectrograph slit 
relative to the major axis of the galaxy
(if we are indeed dealing with disks).\footnote{The slit
orientation on the sky was dictated by the location
of the offset star relative to the LBG---see \S2.1\,.}

We can check on points (b) and (c) by examining
existing high resolution images of these two LBGs.
Q0347$-$383~C5 was part of the WFPC2 sample 
of Giavalisco et al. (1996)\footnote{The galaxy was labeled
0347$-$383--N05 in that paper.}; in Figure 11 we show
the image obtained by combining ten dithered 1800\,s  
exposures through the F702W filter 
(which is a very close match to our {$\cal R$) filter)
after `drizzling' onto a master output pixel grid
(Fruchter \& Hook 1998).
Overlaid on the WFPC2 image is the location on the sky
of the 1 arcsec wide ISAAC slit.
Q0347$-$383~C5 exhibits an irregular morphology,
with a knot of intense UV emission and extensions 
of lower surface brightness to the north.
The horizontal arrows in Figure 11 delimit the location
of the [O~III] emission 
which is clearly {\it not\/}¥ coincident 
with the UV light\footnote{In this case we can 
register precisely the ISAAC 2-D spectrum and the 
WFPC2 image because they both include the bright
QSO Q0347$-$383 which is located only 26.7~arcsec
away from the LBG.}.
With hindsight this is not totally unexpected;
some possible causes have been considered in \S4.
Similar differences in the 
spatial distributions of 
H~II regions and stellar UV continuum 
have been observed in nearby starburst galaxies, 
albeit on smaller scales
(e.g. Leitherer et al. 1996). 
Although the spectrograph slit was fortuitously
aligned so as to encompass the extended structure
revealed by the WFPC2 image, one can hardly
interpret such structure as evidence for a rotating disk,
or hazard a guess as to the inclination angle on the sky.
Furthermore, some of the apparent rotation in Figure 9 
may in reality be an instrumental effect caused
by the clumpy structure of the galaxy,
if knots of peak [O~III] emission are centered at different
locations {\it across\/}¥ the slit.

In Figure 12 we have overlaid the ISAAC slit
on a $K$-band image of SSA22a~MD46 obtained with 
NIRC as part of the survey by Shapley et al. (in preparation).
In this case our IR spectrum missed a second knot
of continuum emission which evidently fell outside 
the spectrograph slit; it would be interesting to repeat
the observation with the slit rotated by 90 degrees
so as to probe the kinematics of the gas along the apparent
elongation axis of this galaxy.

In summary, while in neither case do we find conclusive evidence
supporting the hypothesis that the extended [O~III] emission 
traces a rotation curve,
these initial results are nevertheless intriguing.
Looking ahead, it should be possible to investigate
more extensively the velocity and spatial structure
of Lyman break galaxies by combining high resolution
imaging with spatially resolved spectroscopy;
this task will be accomplished most effectively
with near-IR spectrographs 
fed by adaptive optics systems.\\

\subsection{Large-scale Motions}

For 17 LBGs we can compare the redshift of the nebular
emission lines, $z_{\rm H~II}$, listed in column (2)
of Table 4, with the redshift of the interstellar 
absorption lines, $z_{\rm abs}$; 
for a subset of 13 objects we can also include
in the comparison the redshift of the \lya\ emission line,
$z_{\rm Ly\alpha}$.
Values of $z_{\rm abs}$ and $z_{\rm Ly\alpha}$
can be found in columns (5) and (4) of Table 1 respectively.
The former is the mean of all the UV interstellar absorption lines
(generally three or more)
between 1250 and 1700\,\AA\
which could be identified in our LRIS spectra;
we refer the reader to Pettini et al. (2000)
for a list of the strongest transitions.
The latter was measured either by gaussian fitting or from the peak of 
the emission, depending on the shape of the \lya\ emission line.
If, as a working assumption, we adopt the values of 
$z_{\rm H~II}$ as the systemic redshifts of the galaxies,
we can convert $z_{\rm abs}$ and $z_{\rm Ly\alpha}$
to the relative velocities listed respectively in columns 
(5) and (6) of Table 4 and plotted in Figure 13.

These data reveal a clear pattern in the kinematics of the 
interstellar medium of Lyman break galaxies when probed with these
three different tracers. In three quarters of the objects
observed the interstellar absorption lines are blueshifted
relative to the H~II region emission lines, while in all
cases \lya\ emission is redshifted. 
$\Delta{\rm v_{IS~abs}}$ is typically between
$\sim -200$ and $\sim -400$\,km~s$^{-1}$, with a 
median value of $-300$\,km~s$^{-1}$; 
values of $\Delta{\rm v_{Ly\alpha}}$ span a larger range,
from $\sim +200$ to $\sim +1100$\,km~s$^{-1}$.
A similar effect was already evident in the
small sample of Pettini et al. (1998);
the new data show it to be a characteristic shared
by most Lyman break galaxies.
There is no evidence for any dependence 
of these large velocity offsets on either
the far-UV or optical luminosities of the galaxies
within the $\sim 2$ magnitude interval in 
$M_{1500}$ and $M_{\rm B}$ probed by the 
present study.  

The simplest interpretation---and undoubtedly an 
oversimplification of a more complex physical picture---is 
that we are seeing galactic scale outflows 
driven by the mechanical energy deposited by 
supernovae and stellar winds 
in these actively star forming galaxies.
Presumably the gas seen in absorption
in front of the stars is the approaching part of
an expanding shell of swept-up material which has a very high optical 
depth to \lya\ photons; thus the only \lya\ emission
detectable along our line of sight is from the 
back of the shell, {\it behind\/}¥ the stars,
receding at velocities where no foreground 
absorption takes place. This behavior of the \lya\ line
is commonly seen in local H~II galaxies
(Kunth et al. 1998) and has been modelled 
extensively by Tenorio-Tagle
et al. (1999) among others.
In MS~1512-cB58, where $\Delta{\rm v_{IS~abs}} = -390$\,km~s$^{-1}$,
Pettini et al. (2000) deduced a mass outflow rate 
$\dot{M} \simeq 60\,M_{\odot}$~yr$^{-1}$,
comparable to the star formation rate
SFR\,$ \simeq 20\,M_{\odot}$~yr$^{-1}$ (Table 2).
Since both the value of SFR and of
$\Delta{\rm v_{IS~abs}}$ 
in MS~1512-cB58 are typical of the present
sample of LBGs, it would appear
that luminous Lyman break
galaxies are {\it generally\/}¥ the sites of powerful
superwinds involving a
mass in baryons comparable
to that being turned into stars.

In this and many other respects the properties of
the superwinds we see at $z \simeq 3$
are very similar to those observed
in nearby starburst galaxies and reviewed extensively
by Heckman (e.g. Heckman 2000).  In the local universe
they are found in galaxies with high
rates of star formation per unit area,
$\Sigma_{\ast} \geq 0.1\,M_{\odot}$~yr$^{-1}$~kpc$^{-2}$;
this threshold is exceeded by one order of magnitude
by the LBGs considered here, which have
typical SFR\,$\simeq 40\,M_{\odot}$~yr$^{-1}$
and half-light radii $r_{1/2} = 2.3$\,kpc.
Similar outflow rates and speeds are involved
at high and low redshift; Heckman et al.
(2000) calculate that the kinetic energy required
is $\approx 10$\% of the total kinetic energy supplied by the 
starburst.

Galactic superwinds have several important
astrophysical consequences which have 
already been considered
in the review by Heckman (2000); here we 
limit ourselves to a few comments of particular
significance at high redshift.
First, the data in Figure 13, and the mass outflow rates
they imply, provide a vivid empirical demonstration
of the feedback process required to regulate
star formation in nearly all theoretical models
of galaxy formation (e.g. Efstathiou 2000; Cole et al. 2000).
Second, the results of \S5 show Lyman break galaxies to be 
the most metal-enriched structures at $z \simeq 3$,
apart from QSOs (Pettini 2000).
A significant fraction of these metals will probably
be lost from the galaxies altogether, since the measured values
of $\Delta{\rm v_{IS~abs}}$ are comparable to 
the escape velocities (Heckman 2000; 
Ferrara, Pettini, \& Shchekinov 2000). While it is still
unclear how far this metal-enriched gas will
travel (Ferrara et al. 2000; Aguirre et al. 2001),
there is at least the potential for 
seeding large volumes around the galaxies 
with the products of stellar nucleosynthesis.
Furthermore, if most of the metals carried away by galactic superwinds 
remain in a hot phase, yet to be directly observed, 
this may help solve the puzzle of the missing metals 
at high redshift (Pettini 1998; Pagel 2000).
Third, in any flattened geometry, it is likely 
that the expanding superbubbles 
will `punch holes' through the interstellar medium
in the direction of the vertical pressure gradient.
Such cavities would allow 
Lyman continuum photons to leak into the IGM. 
The common occurrence of superwinds in Lyman break galaxies
may then provide a plausible explanation for the high escape
fraction of ionizing photons suggested by the 
recent results of Steidel et al. (2001).
Thus, the mechanical energy deposited
by the star formation episodes themselves
may well be the key physical process 
ultimately responsible for reionizing of the universe 
at high redshift.\\

\section{Summary and Conclusions}

We have presented the first results of a spectroscopic
survey of Lyman break galaxies in the near infrared,
aimed at detecting nebular emission lines
of [O~II], [O~III] and \Hb\ with the NIRSPEC and 
ISAAC instruments on the Keck and VLT telescopes respectively.
Together with observations from the literature,
we have constructed a sample of data for 19 LBGs,
the largest considered so far.
The galaxies are drawn from the bright
end of the luminosity function, from 
$\sim L^{\ast}$ to $\sim 4\,L^{\ast}$.
Their near-IR spectra have been analysed to investigate
the star formation rates, dust obscuration, oxygen abundances,
and kinematics of the normal galaxy population
at $z \simeq 3$. The main results of this work are as follows.

(1) The LBGs observed form a very uniform sample in their
near-IR properties. The spectra are dominated by the emission lines,
and the continuum is detected in only two objects, one 
of which---West~MMD11---has an unusually red optical-to-infrared
color with (${\cal R} - K_{\rm AB}$)\,$= 2.72$\,.
In all cases [O~III] is stronger than \Hb\ and [O~II].
The line widths span a relatively small range, with 
values of the one dimensional velocity dispersion 
$\sigma$ between 50 and 115\,km~s$^{-1}$.

(2) The star formation rates deduced from the luminosity
of the \Hb\ emission line agree within a factor of $\sim 2$
with the values implied by the continuum luminosity
at 1500\,\AA\ {\it before any corrections for dust extinction
are applied\/}¥. There is no trend in the present sample
for the former to be larger than the latter, as may have been expected
from the shape of all reddening curves which rise from the optical
to the UV. Evidently, any such differential extinction
must be small compared with the uncertainties in calibrating
these two different measures of the SFR.
This conclusion is in agreement with the results of similar 
recent studies of UV-selected star-forming galaxies at $z \simlt 1$
and contradicts the commonly held view that the Balmer
lines are more reliable star formation indicators than the UV 
continuum---from our sample one would obtain essentially the same
star formation rate density using either method.

(3) In five cases (four new ones and one previously published)
we attempted to deduce values of the abundance of oxygen
by applying the familiar $R_{23}$ method which has 
been extensively used in local H~II regions.
We found that generally there remains a significant 
uncertainty, by up to 1\,dex, in the value of (O/H)
because of the double-valued nature of the 
$R_{23}$ calibrator. Thus, in the galaxies observed
oxygen could be as abundant as in the interstellar medium 
near the Sun, or as low as $\sim 1/10$ solar.
While this degeneracy can in principle 
be resolved by measuring the [N~II]/\Ha\ ratio
(and in the one case where this has proved possible
values of (O/H) near the upper end of the range are indicated), 
this option is not normally available for galaxies at $z \simeq 3$
because the relevant lines are redshifted beyond the $K$-band. 

Even so, it is still possible to draw some interesting conclusions.
First, LBGs are definitely more metal-rich than
damped \lya\ systems at the same epoch, which typically have 
metallicities $Z \approx 1/30\,Z_{\odot}$. 
This conclusion is consistent with the view that DLAs are drawn 
preferentially from the faint end of the galaxy 
luminosity function and are not the most actively star 
forming galaxies, as indicated by essentially all 
attempts up to now to detect them via direct imaging.
Second, LBGs do not conform to today's
metallicity-luminosity relation and are overluminous
for their oxygen abundance.
This is probably an indication that they have
relatively low mass-to-light ratios, as also
suggested by their kinematical masses; 
a further possibility is that
the whole (O/H) vs. $M_{\rm B}$ correlation shifts
to lower metallicities at high $z$, when galaxies were
younger.

(4) If the emission line widths reflect the relative
motions of H~II regions within the gravitational potential
of the galaxies, the implied masses are of the order
of $10^{10}\,M_{\odot}$ within half-light
radii of $\sim 2.5$\,kpc. This is likely
to be a lower limit to the total masses of the galaxies
as would be obtained, for example, if we could trace their 
rotation curves. A more serious uncertainty, however,
is the real origin of the velocity dispersions we measure.
We do not see any correlation between $\sigma$ and galaxy luminosity
in either our limited sample nor in an on-going
study by some of us of a much larger 
sample of galaxies at $z \simeq 1$ which span five magnitudes in 
luminosity and yet show very similar line widths to those found here.
In two cases we have found hints of ordered motions
in spatially resolved profiles of the [O~III] lines,
but attempts to use high resolution images to clarify whether they are 
indicative of rotating disks proved to be inconclusive.

(5) In all the galaxies observed we find evidence for
bulk motions of several hundred km~s$^{-1}$ 
from the velocities of the
interstellar absorption lines---which are systematically
blueshifted---and \lya\ emission---which is always 
redshifted---relative to the nebular emission lines.
We interpret this effect as indicative of 
galaxy-wide outflows which appear to be
a common characteristic of galaxies with large rates of star 
formation per unit area at high, as well as low, redshifts.
Such `superwinds' involve comparable amounts of matter 
as is being turned into stars
(the mass outflow rate is of the same order as the star formation rate)
and about 10\% of the total kinetic energy delivered by the starburst.
Furthermore, they have a number of important astrophysical consequences.
They provide self-regulation to the star formation process;
can distribute the products of stellar
nucleosynthesis over large volumes
(the outflow speeds often exceed the escape velocities);
may account for some of the `missing' metals at high redshift;
and may also allow Lyman continuum photons to leak from the 
galaxies into the intergalactic medium, easing the 
problem of how the universe came to be reionized.
On the other hand, the existence of such large velocity
fields within Lyman break galaxies makes it difficult
to measure precisely their systemic redshifts (as
may be required, for example, for detailed clustering studies)
unless more than one marker is available.

(6) Finally, we point out some instrumental developments
which in our view will greatly aid further progress in this field.  
In particular,
we consider that future spectrographs 
making use of adaptive optics and incorporating
multi-object and integral field facilities
will prove to be particularly 
beneficial for the study of high 
redshift galaxies at near-IR wavelengths.\\

We are indebted to the staff of the Paranal and Keck
observatories for their expert assistance with the observations.
We are especially grateful to Chip Kobulnicky for generously
providing the
low redshift data shown in Figure 7, for communicating 
results in advance of publication, and for valuable
comments which improved the paper.
We acknowledge helpful conversations with
Stephane Charlot, Bernard Pagel, and Roberto Terlevich.
C.C.S. and K.L.A. have been supported by grants 
AST~95-96229 and AST~00-70773 from the US 
National Science Foundation and by the David and 
Lucile Packard Foundation.

\newpage


{}

\newpage

%
%


\begin{figure}
\figurenum{0}
\psfig{figure=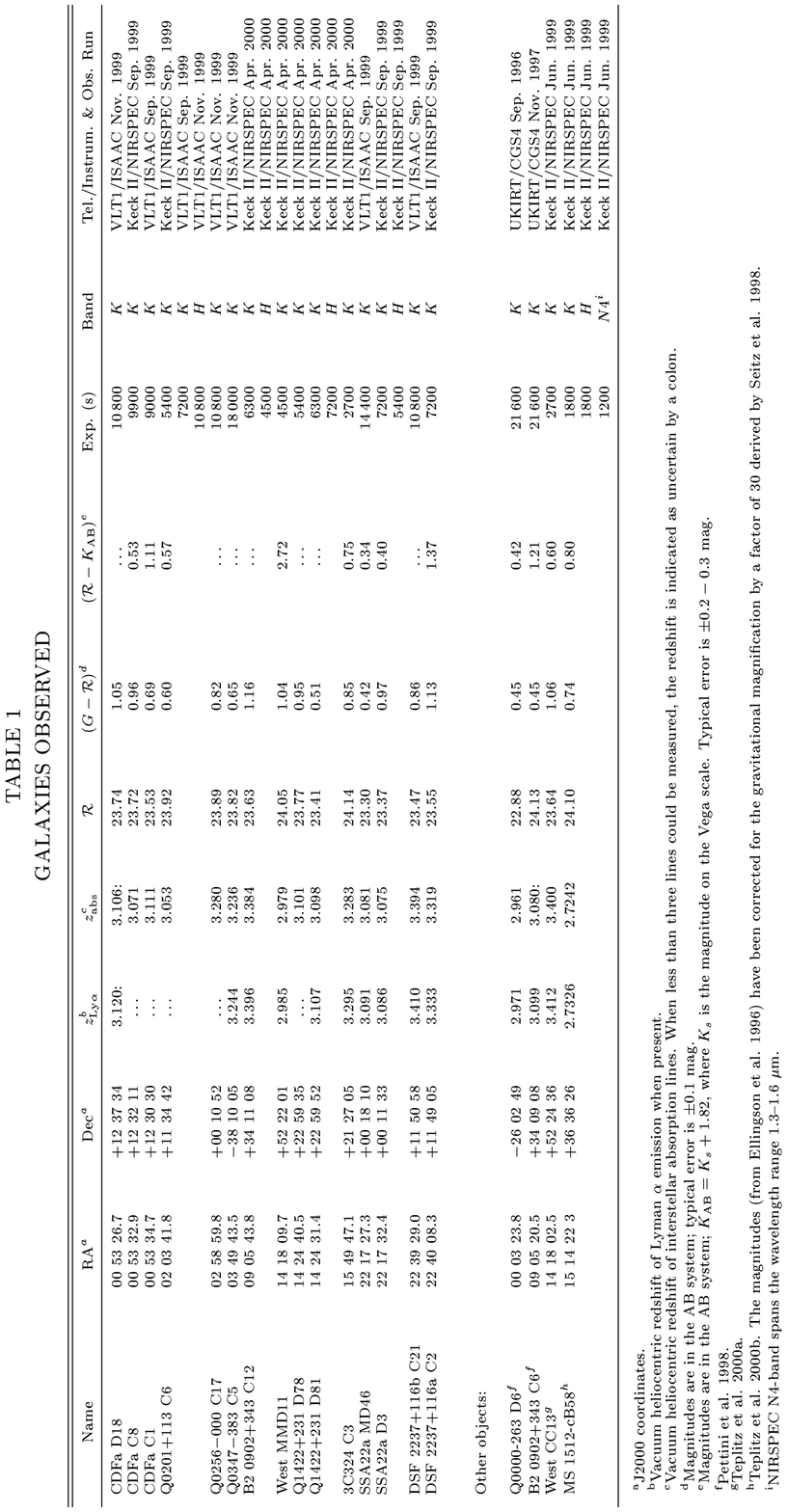,width=170mm,angle=180}
\end{figure}


\begin{figure}
\figurenum{0}
\hspace*{-3.5cm}
\psfig{figure=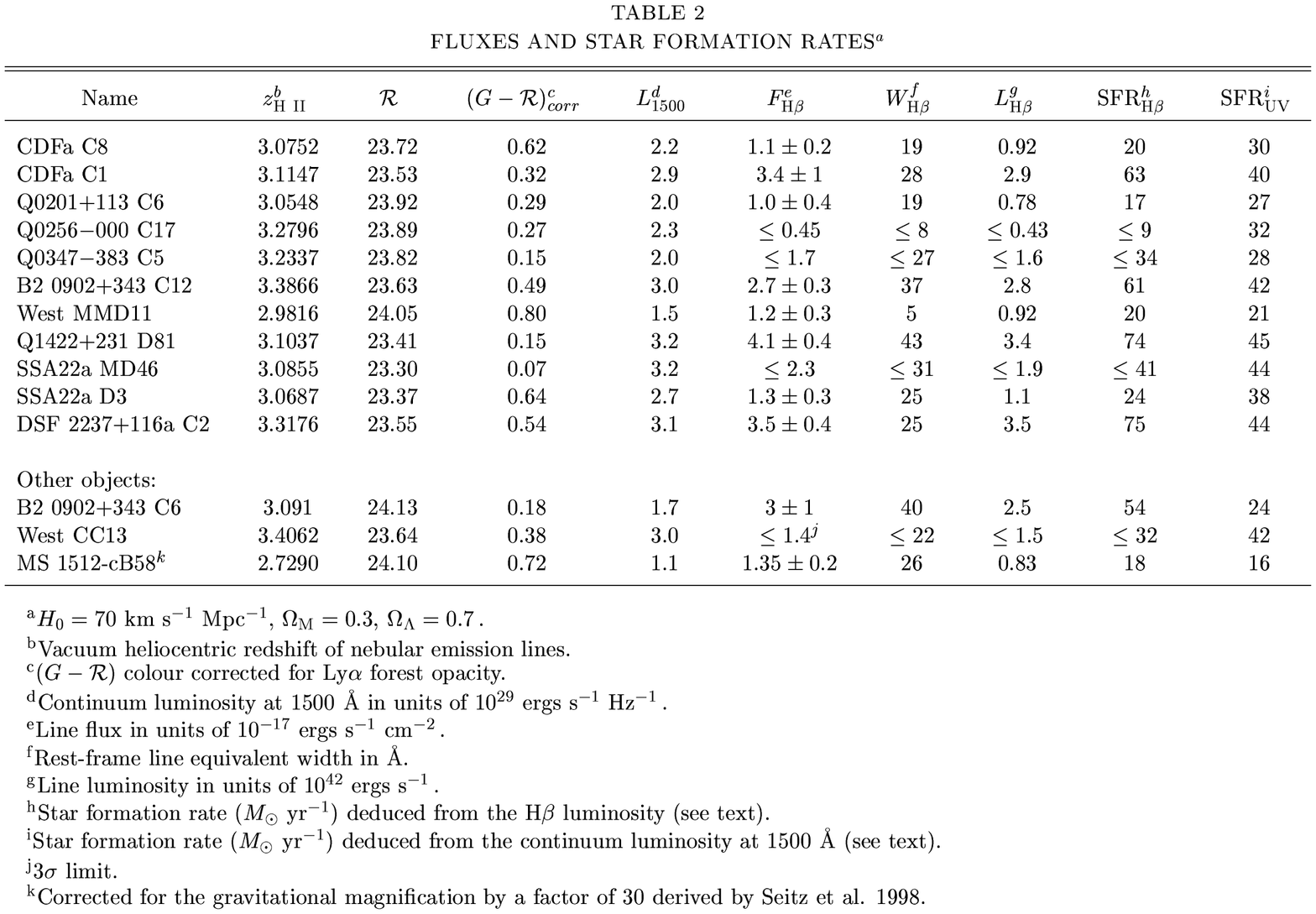,width=160mm,angle=90}
\end{figure}


\begin{figure}
\figurenum{0}
\hspace*{-3.5cm}
\psfig{figure=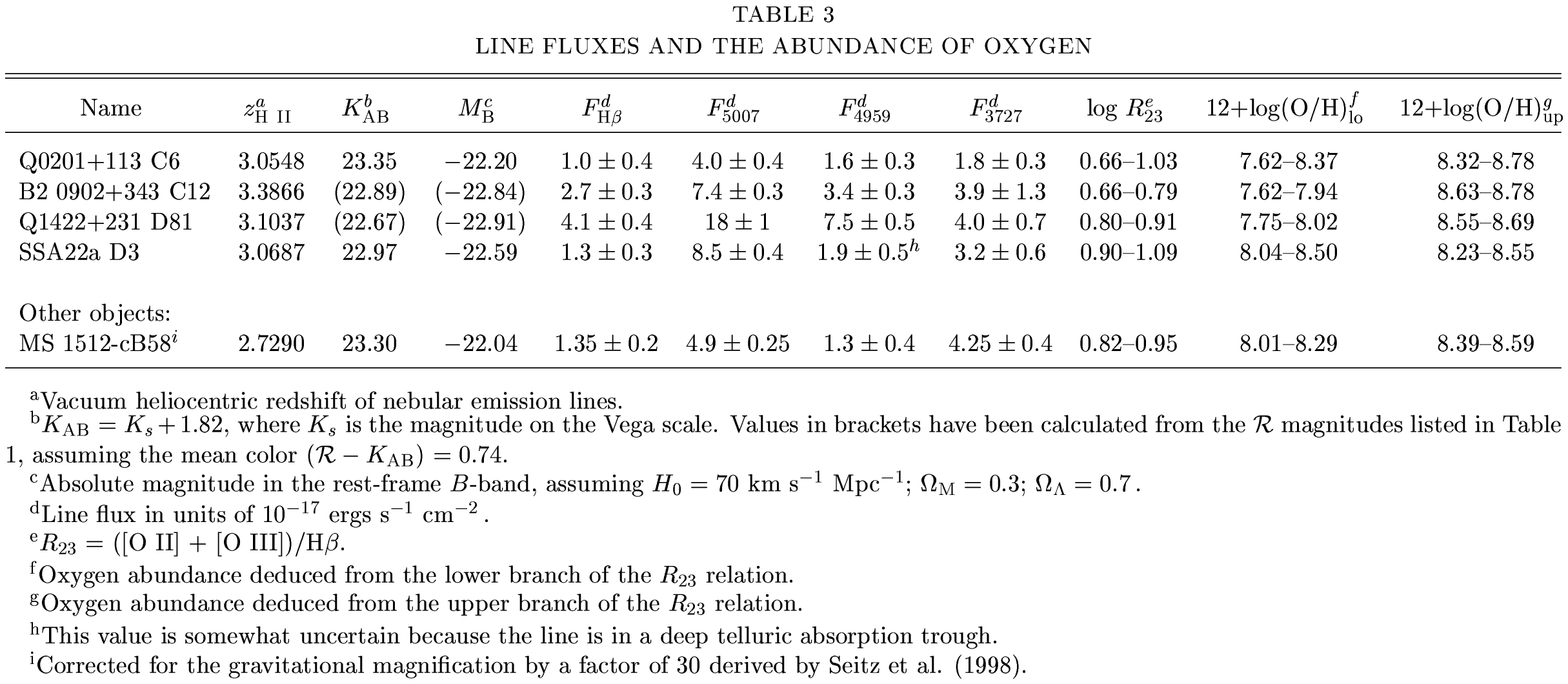,width=160mm,angle=90}
\end{figure}


\begin{figure}
\figurenum{0}
\hspace*{-4.2cm}
\psfig{figure=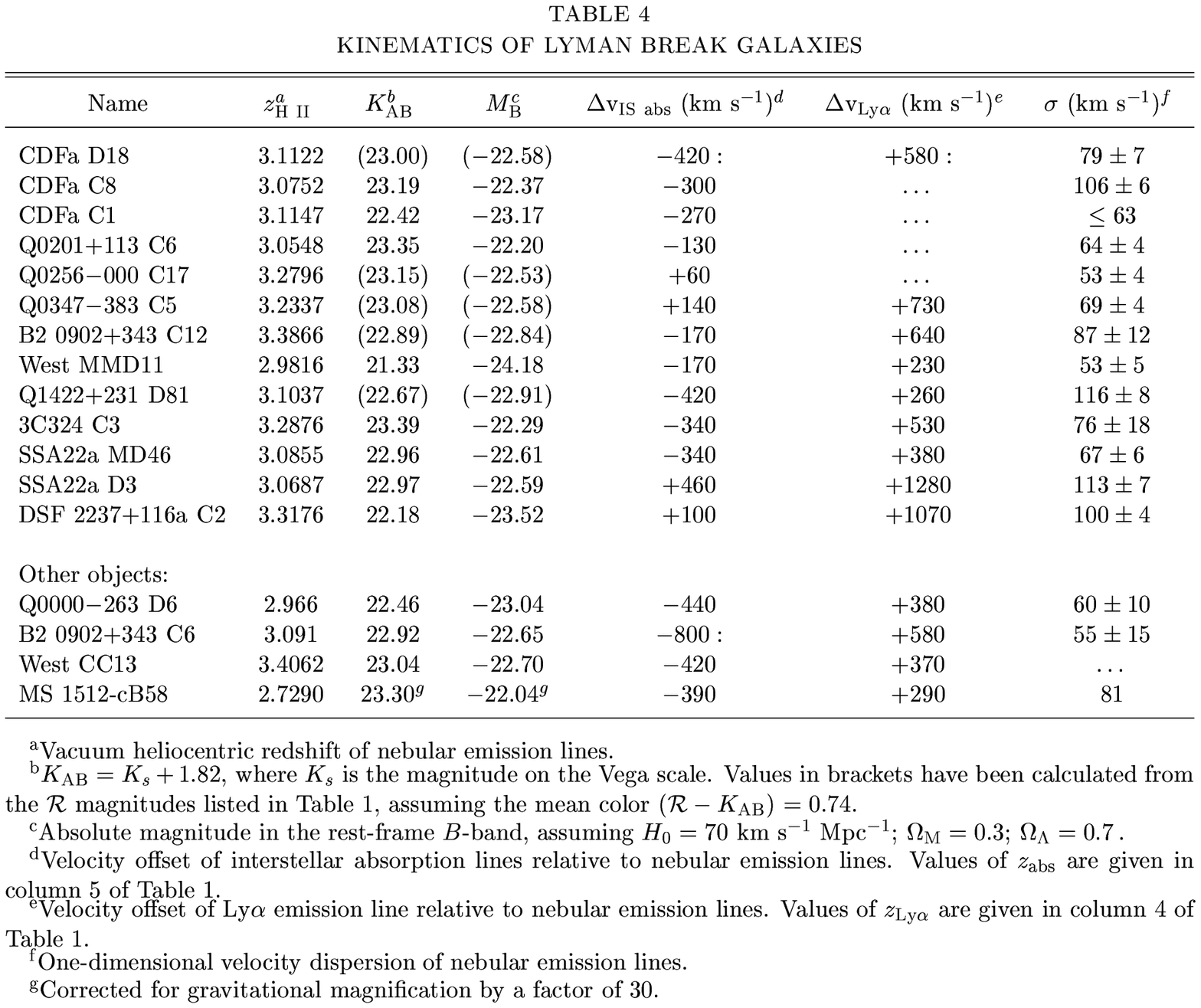,width=160mm,angle=90}
\end{figure}

\newpage

\vfill

%
%

%
%

\begin{figure}
\figurenum{1}
\vspace*{-1.5cm}
\hspace*{-0.50cm}
\psfig{figure=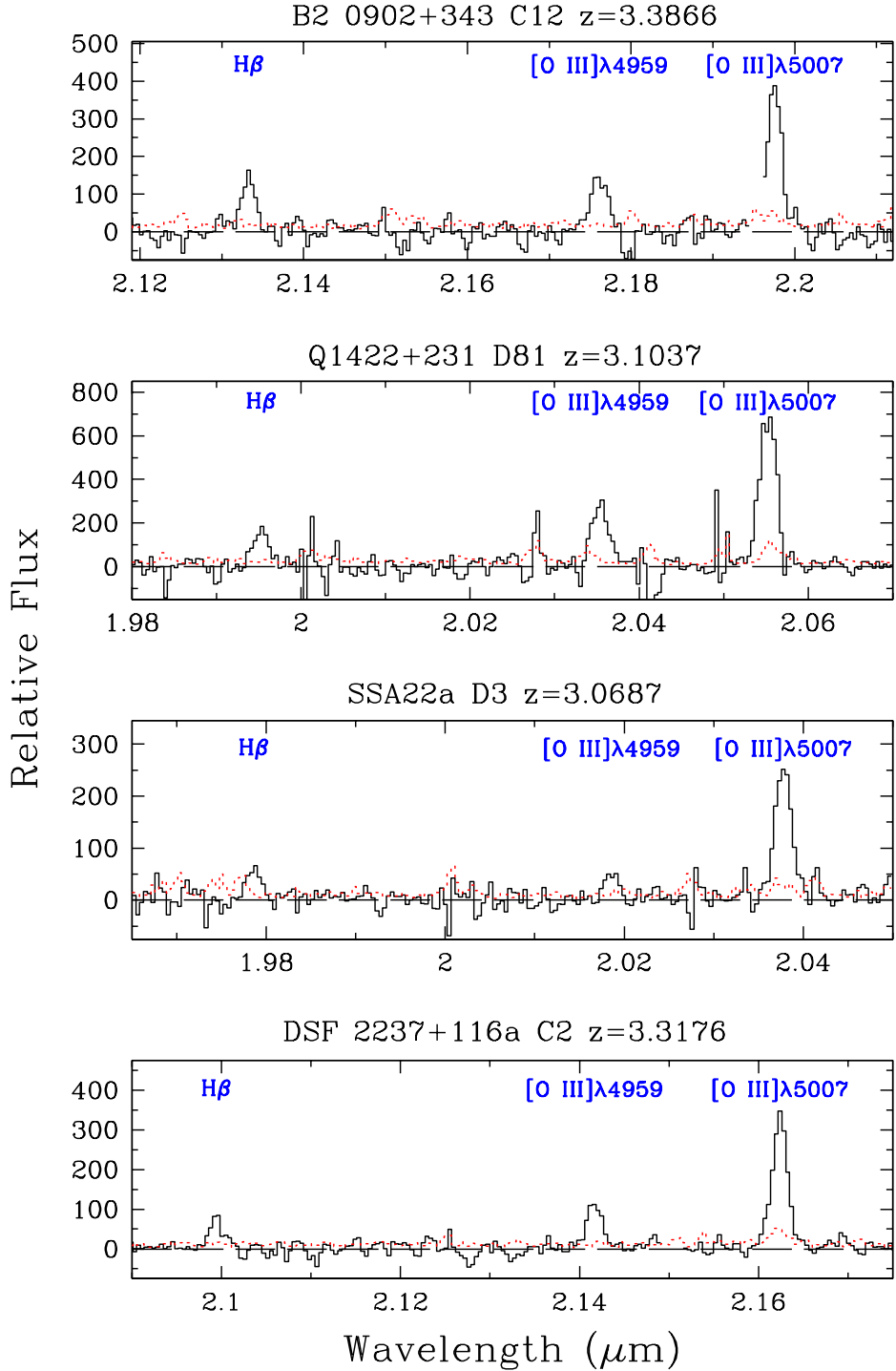,width=165mm}
\vspace{-1.0cm}
\figcaption{Examples of NIRSPEC $K$-band spectra of Lyman break galaxies.
The resolving power is $R \simeq 1500-1750$ sampled with three wavelength 
bins; the exposure time was typically $\sim 2$~hours (see Table 1). 
In each panel the dotted line shows the $1 \sigma$ error spectrum.
Gaps in the spectra correspond to 
invalid data points resulting mostly from poorly subtracted sky lines.
}
\end{figure}

%
%

\begin{figure}
\figurenum{2}
\vspace*{-1.5cm}
\hspace*{-0.50cm}
\psfig{figure=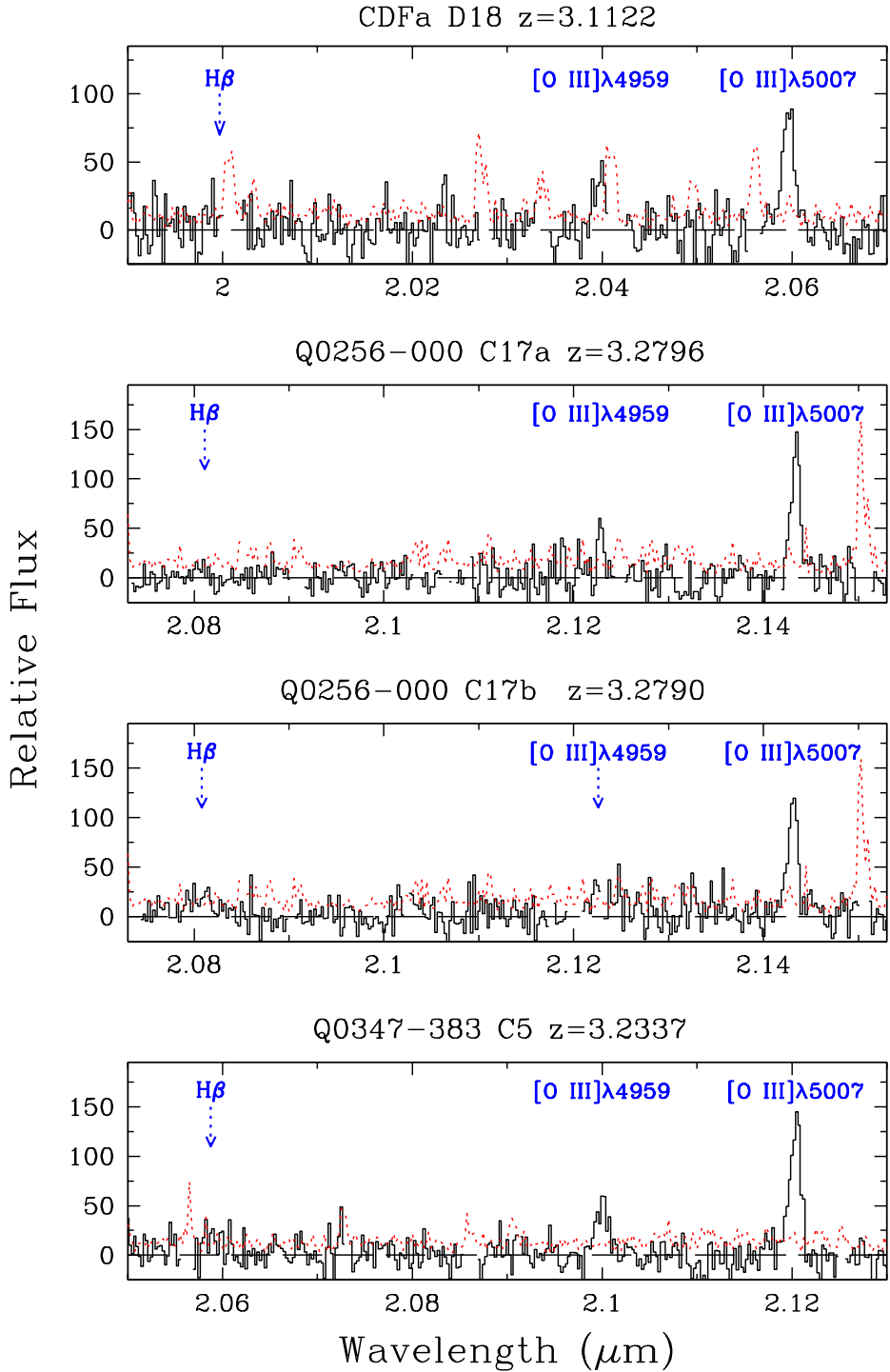,width=165mm}
\vspace{-1.0cm} 
\figcaption{Examples of ISAAC $K$-band spectra of Lyman break galaxies.
The resolving power is $R \simeq 2750$, 
sampled in this plot with three wavelength 
bins (each bin is twice the original pixel size).
The exposure times were three~hours, 
except for Q0347$-$383~C5 for which the exposure time was five hours.
In each panel the dotted line shows the $1 \sigma$ error spectrum;
the expected positions of $un$detected emission lines are indicated by 
downward pointing arrows.  
}
\end{figure}

%
%

\begin{figure}
\figurenum{3}
\vspace*{-3.5cm}
\hspace*{-0.50cm}
\psfig{figure=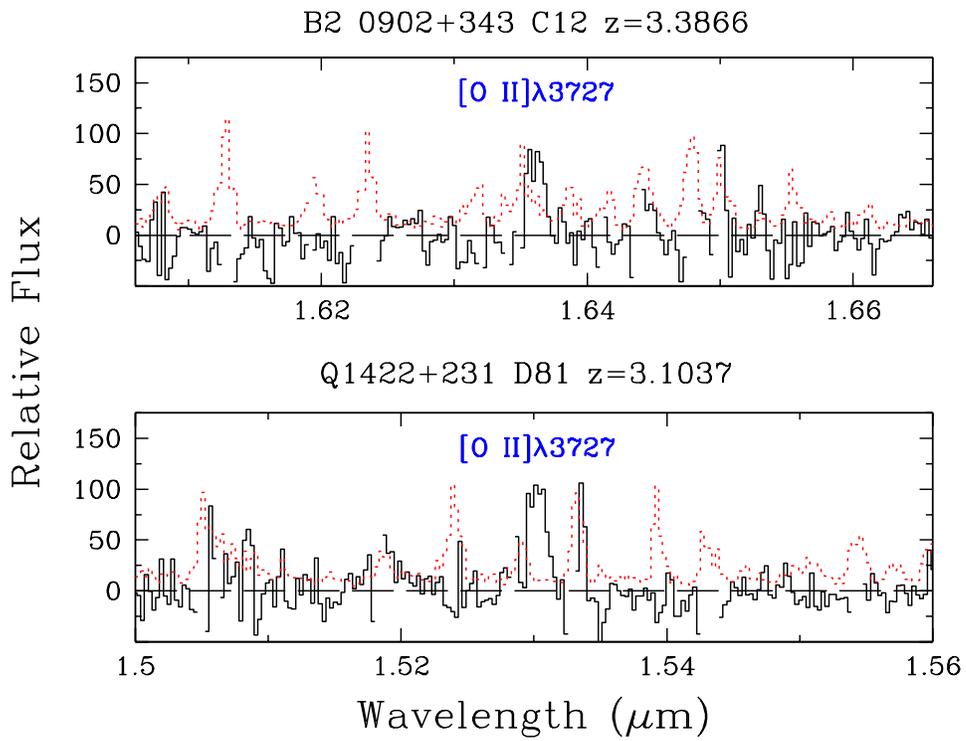,width=165mm}
\vspace{-5.0cm} 
\figcaption{Examples of NIRSPEC $H$-band spectra of Lyman break galaxies.
In each panel the dotted line shows the $1 \sigma$ error spectrum.
}
\end{figure}

%
%

\begin{figure}
\figurenum{4}
\vspace*{-3.5cm}         
\hspace*{-2.5cm}
\psfig{figure=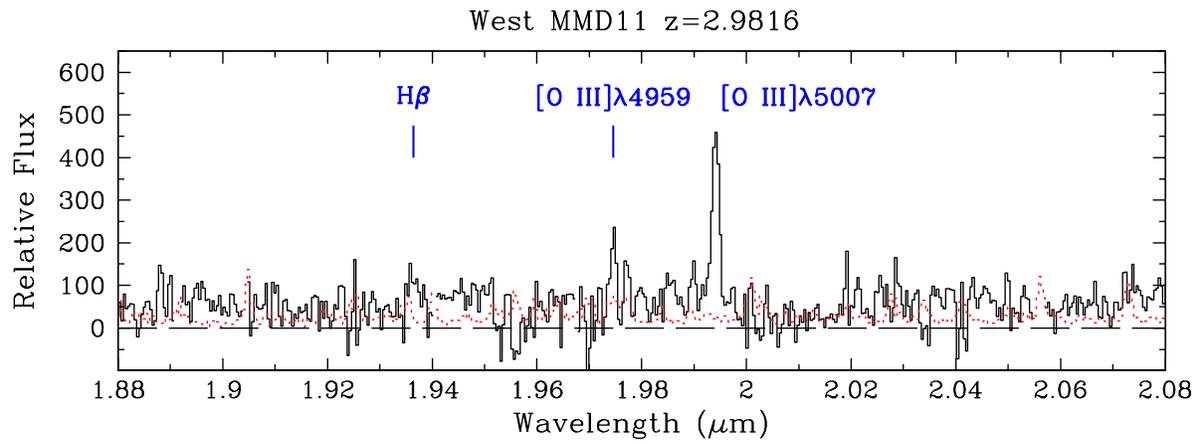,width=155mm,angle=270}
\vspace{-4.0cm}  
\figcaption{NIRSPEC $K$-band spectrum of West MMD11, one of the two 
Lyman break galaxies where we detect a clear continuum signal.
The dotted line shows the $1 \sigma$ error spectrum.  
}
\end{figure}

%
%

\begin{figure}
\figurenum{5}
\vspace*{-2.5cm}         
\hspace*{-4.0cm}
\psfig{figure=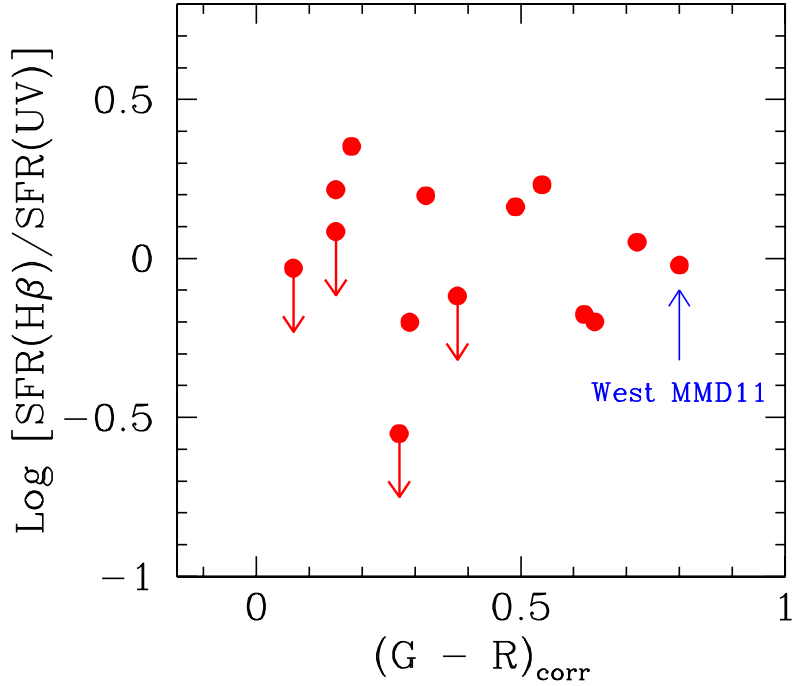,width=185mm,angle=270}
\vspace{-4.0cm}  
\figcaption{Comparison between the values of star formation rate
deduced from the luminosities in the \Hb\ emission line
and in the UV continuum at 1500~\AA.  
The color ($G - {\cal R}$) measures the intrinsic 
UV spectral slope after statistical correction
for the \lya\ forest opacity. Note that in the SCUBA source
West~MMD11, which is also the reddest object in the present sample,
the strength of \Hb\ relative to the UV continuum
is typical of the rest of the sample and 
SFR$_{\rm H\beta} \simeq$\,SFR$_{\rm UV}$.
}
\end{figure}

%
%

\begin{figure}
\figurenum{6}               
\vspace*{-1.75cm}                
\psfig{figure=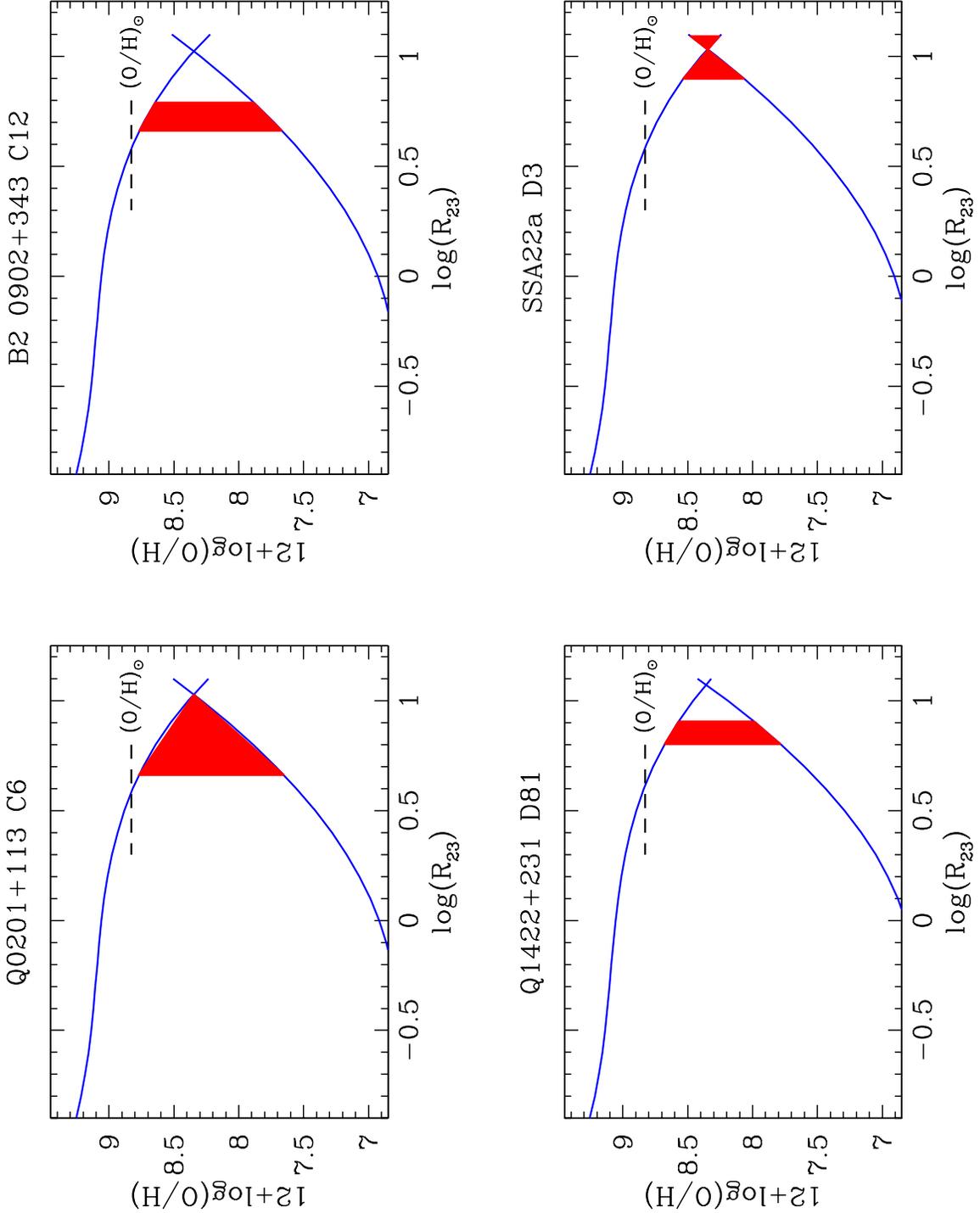,width=165mm}
\vspace{-1.0cm}
\figcaption{Oxygen abundance from the $R_{23}$ = ([O~II]+[O~III])/\Hb\ ratio.
In each panel the continuous lines are the calibration by McGaugh (1991)
for the ionization index $O_{32}$ = [O~III]/[O~II] appropriate to
that object. The shaded area shows the values allowed
by the measured $R_{23}$ and its statistical $1 \sigma$ error.
The broken horizontal line gives for reference the most recent
estimate of the solar abundance 12\,+\,log(O/H) = 8.83 (Grevesse \& 
Sauval 1998).
}
\end{figure}

%
%

\begin{figure}
\figurenum{7}               
\vspace*{-1.75cm}                
\hspace*{-4.75cm}
\psfig{figure=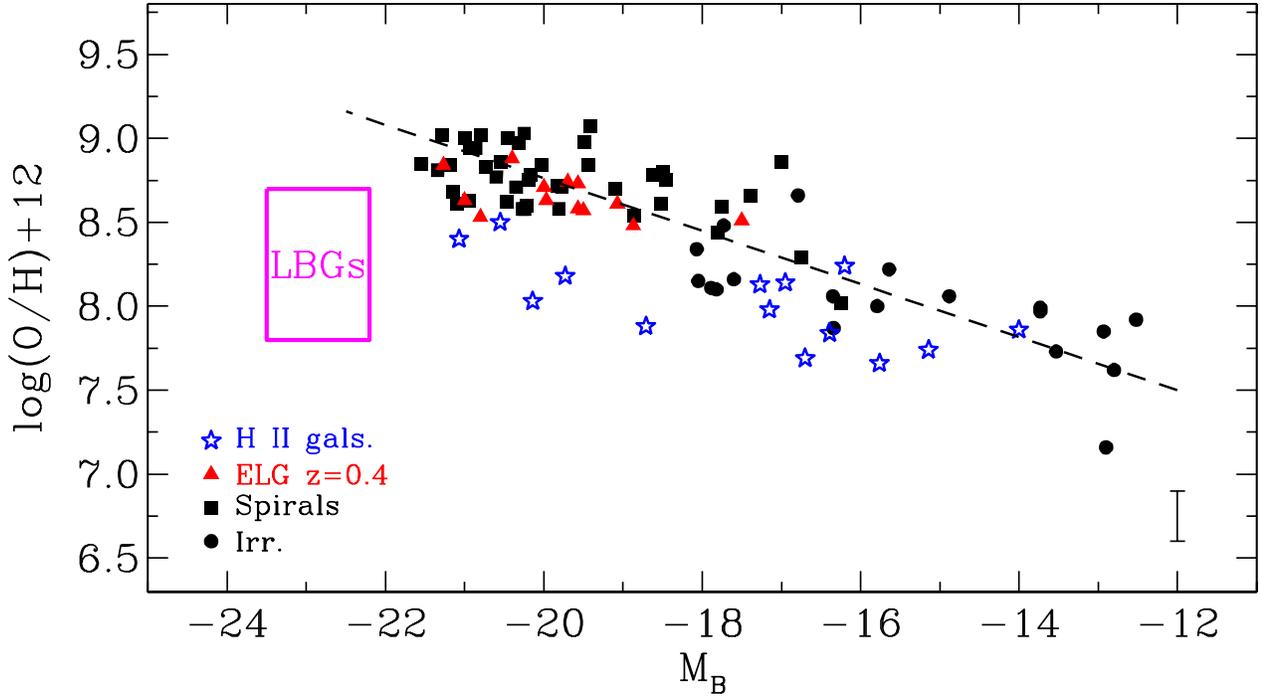,width=190mm,angle=270}
\vspace{-2.0cm}
\figcaption{Metallicity-luminosity relation for local galaxies,
from the compilation by Kobulnicky \& Koo (2000) adjusted to the
$H_0 = 70\,$km~s$^{-1}$~Mpc$^{-1}$, $\Omega_{\rm M} = 0.3$, 
$\Omega_{\Lambda} = 0.7$ cosmology adopted in this paper.
The vertical bar in the bottom right-hand corner
gives an indication of the typical error in log(O/H).
In the Sun 12\,+\,log(O/H) = 8.83 (Grevesse \& Sauval 1998).
The box shows the approximate location of the Lyman break galaxies 
in our sample at a median $z = 3.1$\,. 
Like many local H~II galaxies, 
LBGs are overluminous for their metallicity.
The height of the box results largely from the double-valued nature
of the calibration of (O/H) in terms of the $R_{23}$ index 
(see Figure 6); the one case where the ambiguity can be resolved
(MS~1512-cB58) lies in the upper half of the box.
}
\end{figure}

%
%

\begin{figure}
\figurenum{8}               
\vspace*{-1.75cm}                
\hspace*{-1.75cm}
\psfig{figure=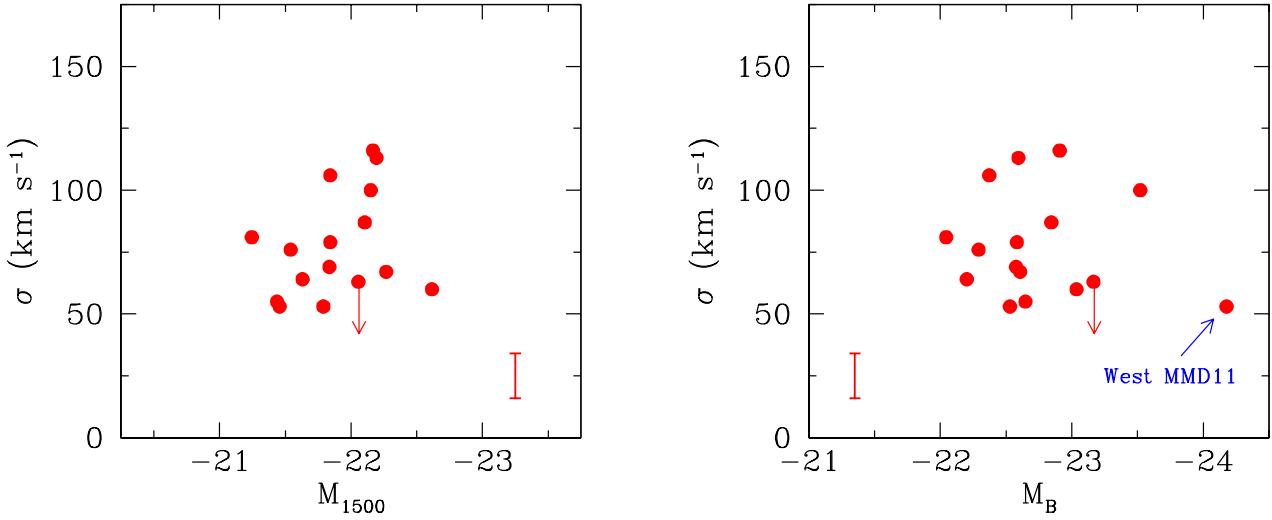,width=140mm,angle=270}
\vspace{-1.0cm}
\figcaption{One dimensional velocity dispersion of nebular 
emission lines in Lyman 
break galaxies as a function of absolute magnitude in the rest-frame
far-UV (left) and $B$-band (right). The vertical bar shows the typical 
error on the measurements of $\sigma$. Curiously,
the SCUBA source West~MMD11, which has the reddest
(${\cal R} - K$) color in the present sample, exhibits
one of the smallest velocity dispersions.
}
\end{figure}

%
%

\begin{figure}
\figurenum{9}               
\vspace*{-2.75cm}                
\hspace*{-1.0cm}
\psfig{figure=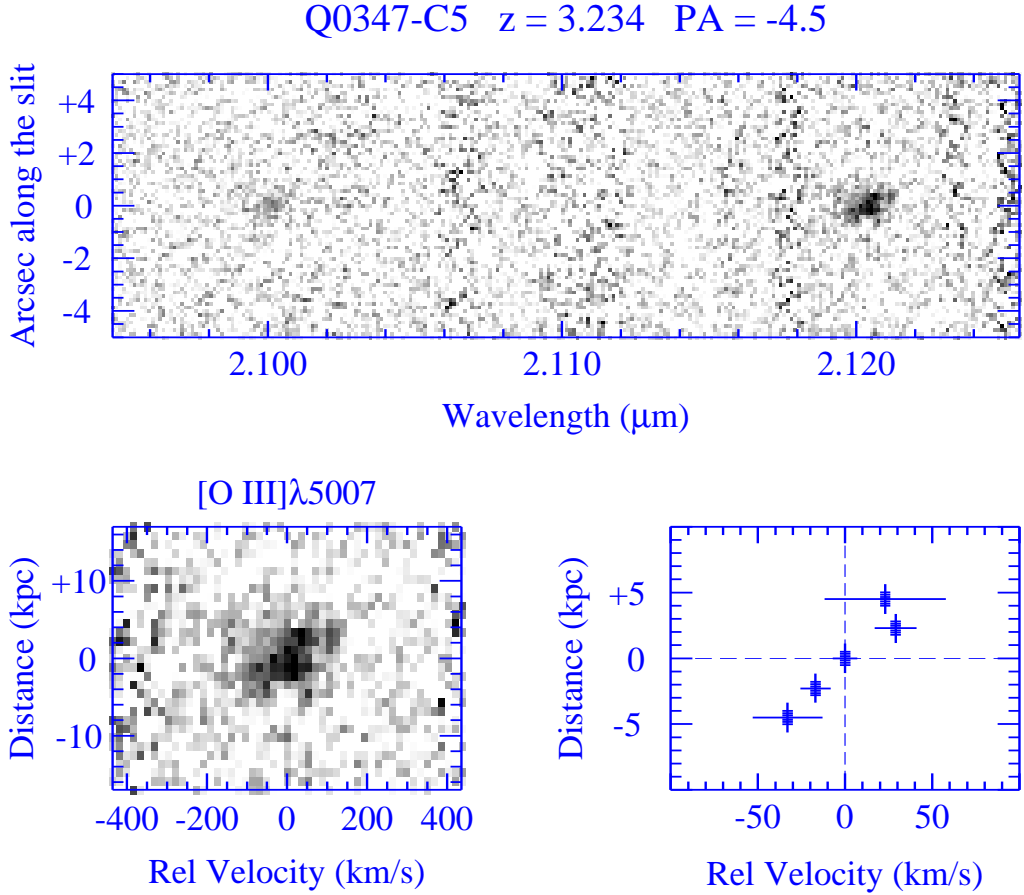,width=140mm,angle=270}
\vspace{+1.0cm}
\figcaption{Spatially resolved emission in ISAAC spectra.
The top panel shows (on a negative scale, so that black is bright)
a portion of the final, stacked
2-D image encompassing the [O~III]~$\lambda\lambda 4959,5007$
doublet obtained with a total integration time of 18\,000\,s. 
The bottom left-hand panel is an enlargement of 
the $\lambda 5007$ line; the projected linear scale
is for our adopted $H_0 = 70$\kms\ Mpc$^{-1}$,
$\Omega_{\rm M} = 0.3$, $\Omega_{\Lambda} = 0.7$,
cosmology. The bottom right-hand panel shows how the 
central wavelength of the emission line varies along the slit.
The spatial resolution of the image is 0.62~arcsec FWHM
(4.7\,kpc).
}
\end{figure}

%
%

\begin{figure}
\figurenum{10}               
\vspace*{-2.75cm}                
\hspace*{-1.0cm}
\psfig{figure=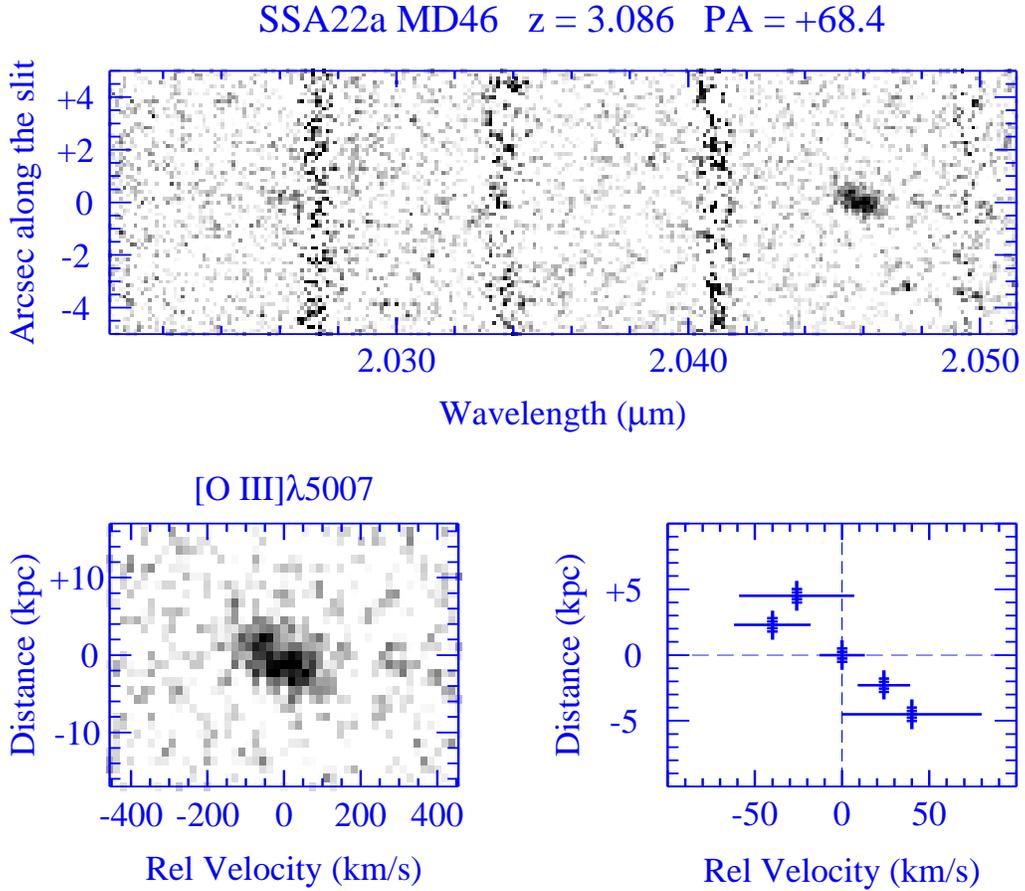,width=140mm,angle=270}
\vspace{+1.0cm}
\figcaption{Spatially resolved emission in ISAAC spectra.
The top panel shows a portion of the final, stacked
2-D image encompassing the [O~III]~$\lambda\lambda 4959,5007$
doublet obtained with an integration time of 14\,400\,s. 
The vertical bands are residuals left from the subtraction
of strong OH sky lines.
The bottom left-hand panel is an enlargement of 
the $\lambda 5007$ line; the bottom right-hand panel shows how the 
central wavelength of the emission line varies along the slit.
The spatial resolution of the image is 0.64~arcsec FWHM
(4.9\,kpc). 
}
\end{figure}        

%
%

\begin{figure}
\figurenum{11}
\vspace*{-3.5cm}
\hspace*{-0.50cm}
\psfig{figure=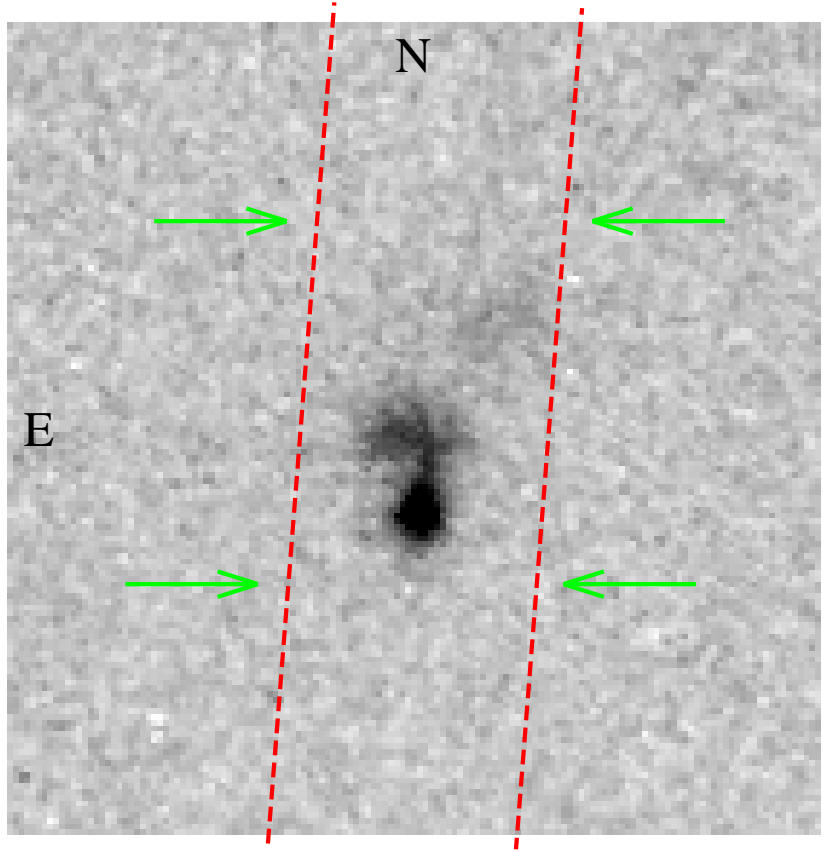,width=165mm}
\figcaption{WFPC2 image of Q0347$-$383~C5 obtained
through the F702W filter with a total exposure time
of 18\,000\,s. The dashed lines indicate the 
edges of the 1 arcsec wide ISAAC slit. The horizontal
arrows mark the limits of the [O~III]~$\lambda 5007$
emission shown in Figure 9, recorded in 0.62~arcsec seeing.
}
\end{figure}

%
%

\begin{figure}
\figurenum{12}
\vspace*{-3.5cm}
\hspace*{-0.50cm}
\psfig{figure=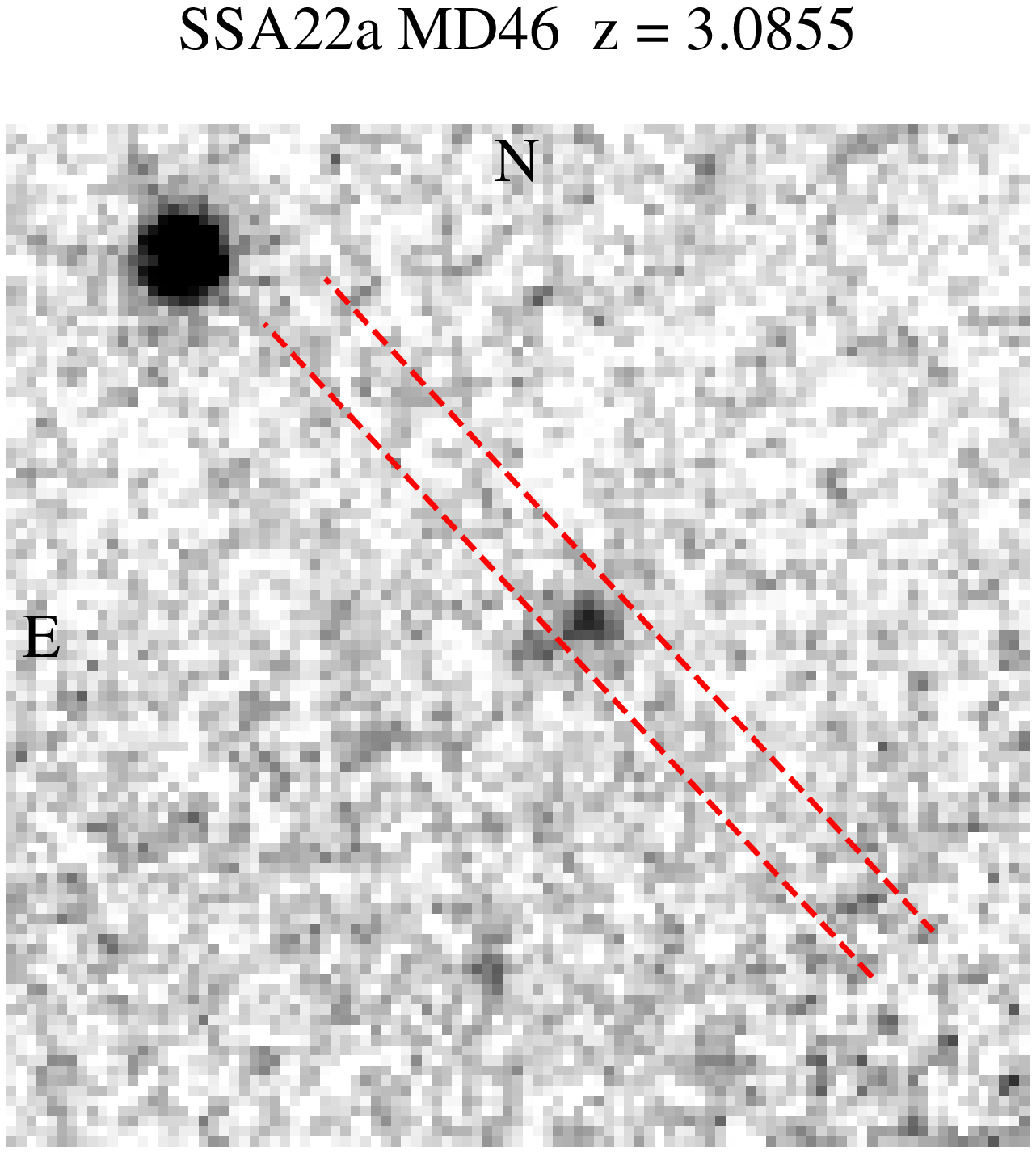,width=165mm}
\figcaption{$K$-band image of SSA22a~MD46 obtained with
NIRC in $\sim 0.5$~arcsec seeing. The total integration time
was 3240\,s made up of 
six sets of exposures, each consisting
of $9 \times 60$\,s frames recorded
in a nine-point dither pattern.
The 1~arcsec wide ISAAC slit is overlaid.
}
\end{figure}

%
%

\begin{figure}
\figurenum{13}               
\vspace*{-2.75cm}                
\hspace*{-3.5cm}
\psfig{figure=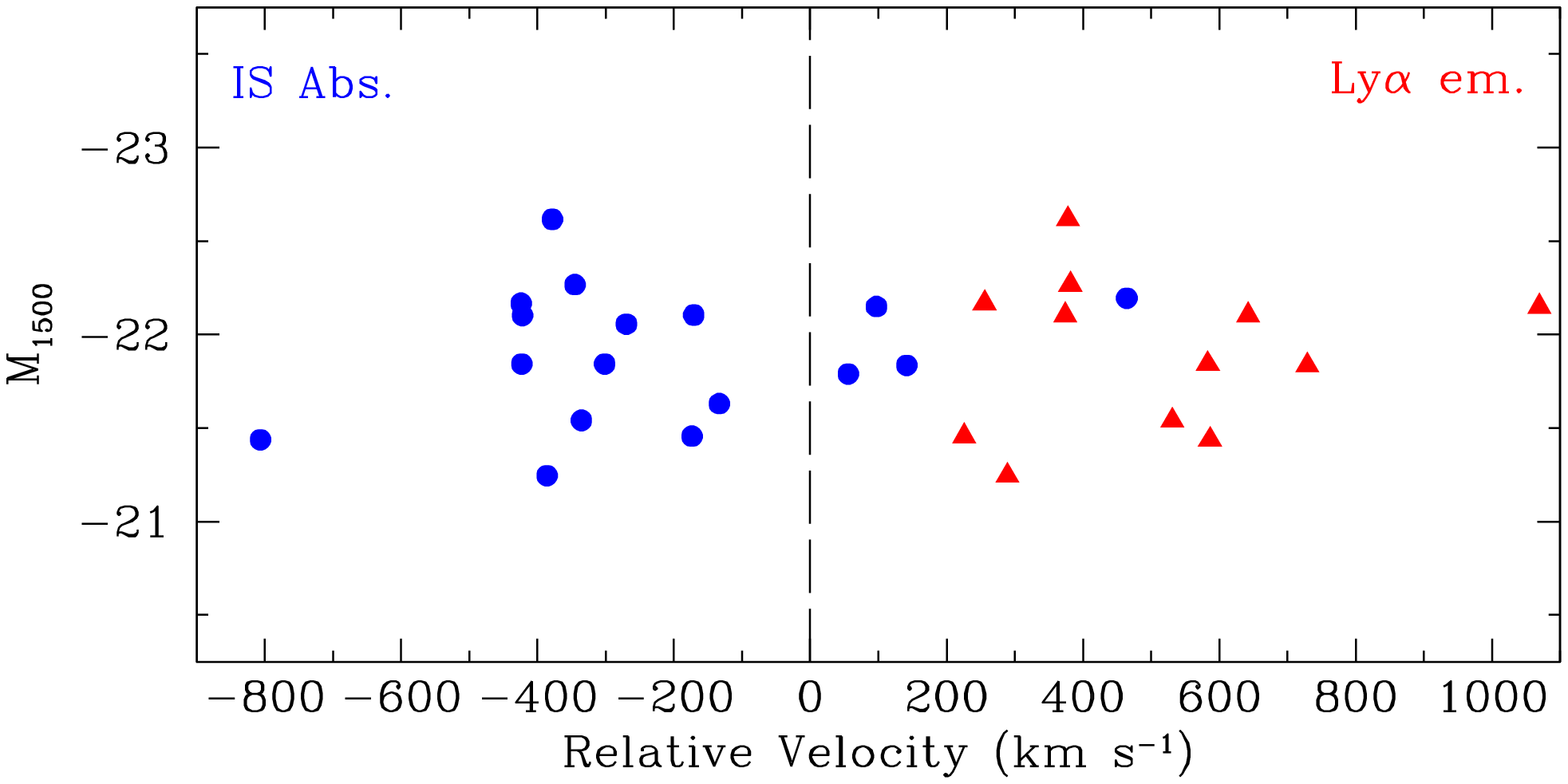,width=175mm,angle=270}
\vspace{-3.0cm}
\figcaption{Velocity offsets in Lyman break galaxies. 
The dots and the triangles correspond respectively to
the values of $\Delta{\rm v_{IS~abs}}$ and 
$\Delta{\rm v_{Ly\alpha}}$ listed in Table 4
and indicate the velocity differences,
relative to [O~III] and \Hb, of the interstellar
absorption lines and of the \lya\ emission line.
Large scale motions of the order of several hundred
km~s$^{-1}$ are indicated by the systematic
tendency for the former to be blueshifted and the latter redshifted
relative to the nebular emission lines.
Plotting these measurements as a function of $M_{\rm B}$
shows a very similar picture---we find no correlation between
values of $\Delta{\rm v}$ and absolute magnitude in either
the far-UV or the optical.
}
\end{figure}


\begin{thebibliography}{}

\bibitem[]{} Adelberger, K.L., \& Steidel, C.C. 2000, ApJ, 544, 218

\bibitem[]{} Aguirre, A., Hernquist, L., Weinberg, D., Katz, N., \&
Gardner, J. 2001, ApJ, submitted (astro-ph/0006345)

\bibitem[]{} Bell, E.F. \& Kennicutt Jr., R.C. 2001, ApJ, in press
(astro-ph/0010340)

\bibitem[]{} Binney, J., \& Tremaine, S. 1987, Galactic Dynamics,
(Princeton:University Press) 

\bibitem[]{} Bunker, A.J., Moustakas, L.A., \& Davis, M. 2000,
ApJ, 531, 95

\bibitem[]{} Bunker, A.J., Warren, S.J., Clements, D.L.,
Williger, G.M., \& Hewett, P.C. 1999, MNRAS, 309, 875

\bibitem[]{} Calzetti, D. 1997, in The Ultraviolet Universe at
at Low and High Redshift: Probing the Progress of Galaxy Evolution,
ed. W.H. Waller, M.N. Fanelli, J.E. Hollis, \& A.C. Danks
AIP Conference Proceedings 408, (New York:Woodbury), 403

\bibitem[]{} Calzetti, D., \& Giavalisco, M. 2000,
Ap \& Sp Sci, in press (astro-ph/0012068)

\bibitem[]{} Chapman, S.C., Scott, D., Steidel, C.C., Borys, C.,
Halpern, M., Morris, S.L., Adelberger, K.L., Dickinson, M.,
Giavalisco, M., \& Pettini, M. 2000, MNRAS, 319, 318

\bibitem[]{} Charlot, S. \& Longhetti, M. 2001, MNRAS, in press
(astro-ph/0101097)

\bibitem[]{} Cole, S., Lacey, C.G., Baugh, C.M., \& Frenk, C.S. 2000, 
MNRAS, 319, 168

\bibitem[]{} Colina, L., Bohlin, R., \& Castelli, F. 1996,
Absolute Flux Calibration Spectrum of Vega, 
STScI Instrument Science Report OSG-CAL-96-01
(Baltimore: STScI)

\bibitem[]{} Cuby, J.G., Barucci, A., de Bergh, C., Emsellem, E.,
Moorwood, A.F.M., Petr, M., Pettini, M., \& Tresse, L. 2000,
Proc. SPIE, 4005, 212

\bibitem[]{} Dickinson, M. 2000, Philos. Trans. R. Soc. Lond. A, 358, 2001

\bibitem[]{} Efstathiou, G. 2000, MNRAS, 317, 697

\bibitem[]{} Ellingson, E., Yee, H.K.C., Bechtold, J., \& Elston, R.
1996, ApJ, 466, L71

\bibitem[]{} Ellison, S.L., Pettini, M., Steidel, C.C. \& Shapley, A.E. 
2001, ApJ, 549, in press.

\bibitem[]{} Esteban, C., Peimbert, M., Torres-Peimbert, S., 
\& Escalante, V. 1998, MNRAS, 295, 401

\bibitem[]{} Ferrara, A., Pettini, M., \& Shchekinov, Y. 2000,
MNRAS, 319, 539


\bibitem[]{} Fruchter, A.S., \& Hook, R.N. 1998, PASP, submitted
(astro-ph/9808087)

\bibitem[]{} Gallego, J., Zamorano, J., Arag\'{on}-Salamanca, A.,
\& Rego, M. 1995, ApJ, 455, L1

\bibitem[]{} Giavalisco, M., Steidel, C.C., \& Macchetto, D. 1996, ApJ, 470,
189

\bibitem[]{} Glazebrook, K., Blake, C., Economou, F., Lilly, S., 
\& Colless, M. 1999, MNRAS, 306, 843

\bibitem[]{} Grevesse, N., \&  Sauval, A.J. 1998, Space Sci Rev, 85, 161

\bibitem[]{} Heckman, T.M. 2000, in ASP Conf. Ser.,
Gas and Galaxy Evolution,
ed. J.E. Hibbard, M.P. Rupen, \& J.H. van Gorkom,
(San Francisco:ASP), in press (astro-ph/0009075)

\bibitem[]{} Heckman, T.M., Lehnert, M., Strickland, D., 
\& Armus, L. 2000, ApJS, 129, 493

\bibitem[]{} Ho, L.C., Filippenko, A.V., \& Sargent, W.L.W. 1997,
ApJ, 487, 579

\bibitem[]{} Kennicutt Jr., R.C. 1998, ARA\&A, 36, 189

\bibitem[]{} Kobulnicky, H.A., Kennicutt Jr., R.C., \& Pizagno, J.L. 1999,
ApJ, 514, 544

\bibitem[]{} Kobulnicky, H.A., \& Koo, D.C. 2000, ApJ, 545, 712
 
\bibitem[]{} Kobulnicky, H.A., \& Zaritsky, D. 1998, ApJ, 511,188

\bibitem[]{} Kulkarni, V.P., Hill, J.M., Schneider, G., Weymann, R.J., 
Storrie-Lombardi, L.J., Rieke, M.J., Thompson, R.I., \& Jannuzi, B.T. 2000,
ApJ, 536, 36

\bibitem[]{} Kulkarni, V.P., Hill, J.M., Schneider, G., Weymann, R.J., 
Storrie-Lombardi, L.J., Rieke, M.J., Thompson, R.I., \& Jannuzi, B.T. 2001,
ApJ, in press (astro-ph/0012140)

\bibitem[]{} Kunth, D., Mas-Hesse, J.M., Terlevich, E., Terlevich, R., 
Lequeux, J., \& Fall, S.M. 1998, A\&A, 334, 11

\bibitem[]{} Lehnert, M.D., \& Heckman, T.M. 1996, ApJ, 462, 651

\bibitem[]{} Leitherer, C. 2000, in 
A Decade of HST Observations, ed. M. Livio, K. S. Noll, \& M. Stiavelli
(Cambridge: CUP), in press

\bibitem[]{} Leitherer, C., Le\~{a}o, J.R., Heckman, T.M., Lennon, D.J.,
Pettini, M., \& Robert, C. 2001, ApJ, 550, in press

\bibitem[]{} Leitherer, C., Schaerer, D., Goldader, J.D.,
Gonz\'{a}lez Delgado, R.M., Robert, C., Kune, D.F., de Mello, D.F., 
Devost, D., \& Heckman, T.M. 1999, ApJS, 123, 3

\bibitem[]{} Leitherer, C., Vacca, W.D., Conti, P.S., Filippenko, A.V.,
Robert, C., \& Sargent, W.L.W. 1996, ApJ, 465, 717


\bibitem[]{} Lowenthal, J., et al. 1997, ApJ, 481, 673

\bibitem[]{} Madau, P. 1995, ApJ, 441, 18


\bibitem[]{} McGaugh, S. 1991, ApJ, 380, 140

\bibitem[]{} McLean, I.S., et al. 1998, Proc. SPIE, 3354, 566

\bibitem[]{} Meyer, D.M., Jura, M., \& Cardelli, J.A. 1998, ApJ, 493, 222

\bibitem[]{} Moorwood, A.F.M., et al. 1999, ESO Messenger, 91, 9

\bibitem[]{} Moorwood, A.F.M., van der Werf, P.P., Cuby, J.G., 
\& Oliva, E. 2000, A\&A, 362, 9

\bibitem[]{} Osterbrock, D.E. 1989, Astrophysics of Gaseous
Nebulae and Active Galactic Nuclei (Mill Valley:University Science Books)

\bibitem[]{} Pagel, B.E.J. 2000, in Galaxies in the Young Universe,
ed. H. Hippelein (Berlin:Springer-Verlag), in press (astro-ph/9911204)

\bibitem[]{} Pagel, B.E.J., Edmunds, M.G., Blackwell, D.E.,
Chun, M.S., \& Smith, G. 1979, MNRAS, 189, 95

\bibitem[]{} Pettini, M. 1999, in Chemical Evolution from Zero to High 
Redshift, ed. J.R. Walsh, \& M.R. Rosa (Berlin:Springer-Verlag), 233

\bibitem[]{} Pettini, M. 2000, Philos. Trans. R. Soc. Lond. A, 358, 2035

\bibitem[]{} Pettini, M., Ellison, S.L., Steidel, C.C. \& Bowen, D.V. 1999,
ApJ, 510, 576

\bibitem[]{} Pettini, M., Kellogg, M., Steidel, C.C., Dickinson, M.,
Adelberger, K.L., \& Giavalisco, M. 1998, ApJ, 508, 539

\bibitem[]{} Pettini, M., Steidel, C.C., Adelberger, K.L., Dickinson, M.,
\& Giavalisco, M. 2000, ApJ, 528, 96

\bibitem[]{} Pilyugin, L.S. 2000, A\&A, 362, 325

\bibitem[]{} Pisano, D.J., Kobulnicky, H.A., Guzm\'{a}n, R.,
Gallego, J., \& Bershady, M.A. 2001, ApJ, submitted.

\bibitem[]{} Prochaska, J.X., Gawiser, E., \& Wolfe, A.M. 2001,
ApJ, in press

\bibitem[]{} Rix, H.W., Guhathahurta, P., Colless, M., \& Ing, K. 1997, 
MNRAS, 285, 779

\bibitem[]{} Rousselot, P., Lidman, C., Cuby, J.G., Moorels, G.,
\& Monnet, G. 2000, A\&A, 354, 1134

\bibitem[]{} Salzer, J.J., Gronwall, C, Lipovetsky, V.A., Kniazev, A.,
Moody, J.W., Boroson, T.A., Thuan, T.X., Izotov, Y.I., Herrero, J.L., \&
Frattare, L.M 2000, AJ, 120, 80

\bibitem[]{} Seitz, S., Saglia, R.P., Bender, R., Hopp, U., Belloni, P., \&
Ziegler, B. 1998, MNRAS, 298, 945

\bibitem[]{} Skillman, E.D., Kennicutt Jr., R.C., \& Hodge, P. 1989, ApJ, 
347, 875

\bibitem[]{} Steidel, C.C. 2000, Proc. SPIE, 4005, 222

\bibitem[]{} Steidel, C.C., Adelberger, K.L., Giavalisco, M.,
Dickinson, M., \& Pettini, M. 1999, ApJ, 519, 1

\bibitem[]{} Steidel, C.C., Giavalisco, M., Pettini, M.,
Dickinson, M., \& Adelberger, K.L. 1996, ApJ, 462, L17

\bibitem[]{} Steidel, C.C., Pettini, M., \& Adelberger, K.L. 2000, ApJ, 546
in press

\bibitem[]{} Sullivan, M., Treyer, M.A., Ellis, R.S., Bridges, T.J., 
Milliard, B., \& Donas, J. 2000, MNRAS, 312, 442

\bibitem[]{} Tenorio-Tagle, G., Silich, S.A., Kunth, D., 
Terlevich, E., \& Terlevich, R. 1999,
MNRAS, 309, 332

\bibitem[]{} Teplitz, H.I., et al. 2000a, ApJ, 542, 18

\bibitem[]{} Teplitz, H.I., et al. 2000b, ApJ, 533, L65

\bibitem[]{} Terlevich, R.J., Denicolo, G., \& Terlevich, E. 2001,
in 

\bibitem[]{} Terlevich, R., Melnick, J., Masegosa, J., Moles, M.,
\& Copetti, M.V.F. 1991, A\&AS, 91, 285

\bibitem[]{} Tresse, L., \& Maddox, S.J. 1998, ApJ, 495, 691

\bibitem[]{} van der Bliek, N.S., Manfroid, J., \& Bouchet, P. 
1996, A\&AS, 119, 547

\bibitem[]{} Vogt, N.P., Phillips, A.C., Faber, S.M., 
Gallego, J., Gronwall, C., Guzm\'{a}n, R., Illingworth, G.D.,
Koo, D.C., \& Lowenthal, J.D. 1997, ApJ, 479, L121

\bibitem[]{} Yan, L., McCarthy, P.J., Freudling, W., Teplitz, H.I., 
Malmumuth, E.M., Weymann, R.J., \& Malkan, M.A. 1999, ApJ, 
519, L47


\end{thebibliography}
\end{document}